# Stellar Population Astrophysics (SPA) with the TNG

## 23 IR elemental abundances of 114 giant stars in 41 Open Clusters

S. Bijavara Seshashayana,[1,2] H. Jönsson,[1] V. D'Orazi,[3,4,5] A. Bragaglia,[6] M. Jian,[7] G. Andreuzzi,[8,9] M. Dal Ponte[4]

[1] Materials Science and Applied Mathematics, Malmö University, SE-205 06 Malmö, Sweden
e-mail: shilpa.bijavara-seshashayana@mau.se
[2] Nordic Optical Telescope, Rambla José Ana Fernández Pérez 7, ES-38711 Breña Baja, Spain
[3] Department of Physics, University of Rome Tor Vergata, via della Ricerca Scientifica 1, 00133 Rome, Italy
[4] INAF-Osservatorio Astronomico di Padova, vicolo dell' Osservatorio 5, 35122 Padova, Italy
[5] Dept. of Astronomy & McDonald Observatory, The University of Texas at Austin, 2515 Speedway, Austin, TX 78712, USA
[6] INAF-Osservatorio di Astrofisica e Scienza dello Spazio di Bologna, via Piero Gobetti 93/3, 40129 Bologna, Italy
[7] Department of Astronomy, Stockholm University, AlbaNova University Center, Roslagstullsbacken 21, 114 21 Stockholm, Sweden
[8] INAF - Osservatorio Astronomico di Roma, via Frascati 33, 00178, Monte Porzio Catone, Italy
[9] Fundacíon Galileo Galilei - INAF, Rambla José Ana Fernández Pérez 7, 38712, Breña Baja, Tenerife, Spain

**ABSTRACT**

*Context.* Open clusters have been extensively used as tracers of Galactic chemical evolution, as their constituent stars possess shared characteristics, including age, Galactocentric radius, metallicity, and chemical composition. By examining the trends of elemental abundances with metallicity, age, and Galactocentric radius, valuable insights can be gained into the distribution and nucleosynthetic origins of chemical elements across the Galactic disc. The infrared domain in particular facilitates the observation of some elemental abundances that can be challenging or impossible to discern in the optical, for example K and F.

*Aims.* The objective of this study is to derive the stellar parameters and elemental abundances of up to 23 elements in 114 stars spanning 41 open clusters using high-resolution infrared spectroscopy. In addition, the present study aims to examine the chemical evolution of the Galactic disc. This is achieved by investigating radial abundance gradients, variations in abundance between clusters, and the dependence of chemical abundances on cluster age.

*Methods.* The spectra utilised in this study were obtained with the high-resolution near-infrared GIANO-B spectrograph at the Telescopio Nazionale Galileo. The derivation of stellar parameters and chemical abundances was achieved by employing the Python version of Spectroscopy Made Easy. In the H-band region, a combination of atomic and molecular features was utilised to constrain the stellar parameters, including OH, CN, CO molecular lines, and Mg I, Si I, Ti I, Ti II, C I and Fe I atomic lines.

*Results.* Abundances for up to 23 elements, C, N, F, Na, Mg, Al, Si, S, K, Ca, Ti, V, Cr, Mn, Fe, Co, Ni, Cu, Zn, Y, Ce, Nd, and Yb were derived and compared with available literature values where possible. Non-local thermodynamic equilibrium analysis was utilised for the elements C, Na, Mg, Al, Si, S, K, Ca, Ti, Mn, Fe and Cu. For each element, Galactic trends were examined by analysing both [X/Fe] and [X/H] as functions of [Fe/H], stellar age, and Galactocentric radius. In particular, the radial abundance gradient of Ytterbium is presented for the first time, thereby extending the observational constraints on heavy neutron-capture elements.

*Conclusions.* Radial abundance gradients for a wide range of elements in the Galactic disc are found, with [X/Fe] slopes ranging from −0.061 to +0.065 dex/kpc. The observed gradients are consistent with an inside-out formation scenario for the Galactic disc, wherein chemical enrichment proceeds from the inner to the outer regions over time. The observed [X/Fe] trends across multiple nucleosynthetic groups, including α-elements, odd-Z elements, iron-peak elements, and neutron-capture elements such as Y, Ce, Nd, and Yb reflect the diverse production sites and timescales associated with each group. In particular, the positive [Zn/H] and [Zn/Fe] gradients suggest a distinctive nucleosynthetic origin for Zn, possibly linked to metallicity-dependent yields. The positive gradient in [Yb/H] (0.065 ± 0.031 dex/kpc) provides significant new constraints on neutron-capture enrichment processes and the chemical evolution of the Galactic disc.

**Key words.** stars: abundances − open clusters and associations: general − Galaxy: evolution − Galaxy: disc

## 1. Introduction

The derivation of chemical abundances of different elements is of great importance in tracing the chemical evolution of our Galaxy. Moreover, in order to achieve a comprehensive understanding of this process of Galactochemical evolution, it is essential to analyse a diverse range of stars, with different ages, Galactocentric radii and metallicities. In recent years, several studies have been conducted in the optical range, which have contributed to the derivation of chemical abundances. However, the study of chemical evolution has also made significant progress in the near-infrared (NIR) spectral range, with the inclusion of stars in dusty regions that would otherwise have obscured optical photons. The study of $\alpha$ and neutron capture elements in the NIR region (see e.g. Montelius et al. 2022, Nandakumar et al. 2024a and references therein) has facilitated our understanding of Galactic chemical evolution. Large-scale spectroscopic surveys, such as Apache Point Observatory Galactic Evolution Experiment (APOGEE, Majewski et al. 2017), have also been instrumental in this regard, providing high-resolution near-infrared (NIR) spectra and abundances of several key elements across many major Galactic components.

The advancement of our comprehension of stellar – as well as Galactic – structure and evolution has been significantly enhanced by the insights gained from the study of star clusters. Among the various types of star clusters, open clusters (OCs) play a crucial role in understanding the characteristics of the low-$\alpha$ disc within galaxies. These clusters are comprised of stars that formed almost simultaneously, from a single molecular cloud, and therefore exhibit the same chemical composition (Spina et al. 2021 and referenes therein). Hence, by examining only a small fraction of their members, it is possible to gain a comprehensive understanding of the cluster's chemical composition. This makes it relatively straightforward to gather data from extensive OC samples. Another advantage of OCs is that their ages and distances can be determined with good precision (e.g., Bragaglia & Tosi 2006; Bossini et al. 2019; Cantat-Gaudin et al. 2020). Consequently, they represent the optimal choice for determining the abundances of diverse elements in stars situated at varying distances from the Galactic center, in disparate azimuthal directions, and at disparate ages.

Despite the comprehensive spectroscopic surveys such as Gaia−European Southern Observatory (Gaia−ESO), APOGEE, and GALactic Archaeology with HERMES (GALAH) (see, e.g., Randich et al. 2022; Donor et al. 2020; Spina et al. 2021) having sub-projects geared towards OCs, the accuracy and precision of the abundances from these industrial analysis pipelines still leave room for dedicated, smaller-scale projects based on more classical analysis methods, that can explore OCs in greater depth. One such study is Open Cluster Chemical Abundances from Spanish Observatories (OCCASO) (Carbajo-Hijarrubia et al. 2024), in which 36 OCs were analyzed using different spectrographs with R > 60000. Another such project is One Star to Tag Them All (OSTTA), which employed the FIES spectrograph at the Nordic Optical Telescope (NOT) (R = 65,000), determining chemical abundances for about 20 poorly studied OCs (Carrera et al. 2022), and has data for more OCs obtained with UVES at the ESO VLT.

The Stellar Populations Astrophysics (SPA) project at the Italian Telescopio Nazionale Galileo (TNG), which employed the HARPS-N and GIANO-B echelle spectrographs, is aligned with these smaller-scale initiatives. The present study is a component of the SPA-OC series, which is designed to derive stellar parameters and abundances of stars in OCs. By comparing the metallicity, age, and Galactocentric distance of a given cluster with the abundance of different elements, one can construct a comprehensive picture of the origin and evolution of these elements across the Galactic disc. A number of other works have been published as part of the SPA-OC series, which analyse giant and main sequence stars: Frasca et al. 2019; D'Orazi et al. 2020; Casali et al. 2020; Alonso-Santiago et al. 2021; Zhang et al. 2021, 2022; Bijavara Seshashayana et al. 2024a; Dal Ponte et al. 2025. Dal Ponte et al. (2025) used optical spectra of 95 stars in 33 open clusters to determine NLTE stellar parameters and abundances but in contrast, we focus exclusively on NIR spectra.

In the present study, the IR abundances of 23 elements for 114 red clump and giant stars in 41 open clusters are presented. These abundances were determined using spectra from the high-resolution NIR spectrograph GIANO-B. This analysis establishes one of the most extensive and homogeneous databases of its kind among "non-indsutial" projects to date. For example, this project is the first to determine elemental abundances in the IR for several clusters, including NGC 7086, UBC 60, UBC 141, UBC 169, UBC 170, and UBC 577. The dataset presented here has a wide age range, from 40 Myr up to 8.3 Gyr, and covers Galactocentric distances ($R_{gc}$) from 6.5 to 11.5 kpc. This enhanced SPA-sample facilitates investigation of the chemical evolution of the Galactic disc with unprecedented detail in the near-infrared, across a wide array of elements: C, N, F, Na, Mg, Al, Si, S, K, Ca, Ti, V, Cr, Mn, Fe, Co, Ni, Cu, Zn, Y, Ce, Nd, and Yb.

## 2. Observations

The data were collected with the 3.5-meter Telescopio Nazionale Galileo (TNG) located at the Observatorio del Roque de los Muchachos on La Palma (Canary Islands, Spain). The telescope has one high-resolution optical spectrometer, the High Accuracy Radial Velocity Planet Searcher for the Northern Hemisphere (HARPS−N) with $R = 115\,000$, $\lambda\lambda = 3800 - 6900$Å, (Cosentino et al. 2014), and one high-resolution NIR spectrometer, GIANO-B, with $R = 50\,000$, $\lambda\lambda = 0.97 - 2.5\mu m$. These instruments were used at the same time in GIARPS (GIANO-B + HARPS-N) mode (Oliva et al. 2012a,b; Origlia 2014; Claudi et al. 2017) during most SPA-observations, but this work pertains exclusively to analysis of the NIR portion. We refer to Dal Ponte et al. 2025 and references therein for the analysis of the optical spectra. The observations were conducted between July 2018 and April 2023. The GOFIO software was used to reduce the spectra (Rainer et al. 2018). This process encompasses the removal of the effects of defective detector pixels, the subtraction of the sky and dark frames, the application of flat-fielding, the optimization of spectrum extraction, and the calibration of wavelengths.

The later data reduction steps are done in the same way as presented in Jian et al. (2025, in review; section 2.1). It includes continuum normalisation and telluric absorption correction. We used the TelFit package to find the best-fit synthetic telluric spectra for each order, and remove those feature by dividing the observed spectra with the telluric synthetic ones. Certain wavelength regions (13530–14350 Å and 18020–19400 Å) are heavily affected by telluric features and they are excluded in the following analysis.

Table A.1 presents a comprehensive overview of the characteristics of the stellar clusters in question. This includes their respective names, the number of stars observed within each cluster, their ages, $R_{gc}$, and the extinction.

## 3. Analysis

The determination of stellar parameters and elemental abundances was achieved through fitting of synthetic spectra to observed spectra, employing the $\chi^2$-minimization technique within predefined regions, using the Python version of Spectroscopy Made Easy (PySME v.0.4.188) (Piskunov & Valenti 2017; Wehrhahn et al. 2023). The selection of a relevant stellar atmosphere model was achieved through the interpolation in a grid of 1D Model Atmospheres in a Radiative and Convective Scheme (MARCS, Gustafsson et al. 2008).

NLTE corrections were applied to atomic spectral lines for C, Na, Mg, Al, Si, S, Ca, K, Ti, Mn, Fe, and Cu (Amarsi et al. 2020; Mallinson et al. 2024; Caliskan et al. 2025).

The atomic data employed in this study is the same as used in (Montelius et al. 2022; Nandakumar et al. 2023a,b, 2024a), so we refer to those papers for a more in-depth description. In short, this line list is based on data from the VALD database (Piskunov et al. 1995; Kupka et al. 2000; Ryabchikova et al. 2015), with astrophysical adjustment of the $\log(gf)$ values of several atomic lines. A listing of the atomic data used in the present work is shown in Table B.1.

### 3.1. Stellar parameters

The initial step in an abundance analysis of stellar spectra entails the determination of stellar parameters, including effective temperature ($T_{eff}$), surface gravity ($\log g$), metallicity ([Fe/H]), microturbulence ($v_{mic}$), and macroturbulence ($v_{mac}$). Accurate parameter determination is paramount for ascertaining the abundances of specific elements. The method presented in this paper is applicable to giants with $3500 < T_{eff} < 5500$ K. The present study's methodology has been devised through the consideration of analogous methodologies with Nandakumar et al. (2023a) and Bijavara Seshashayana et al. (2024b).

The values for $T_{eff}$, [Fe/H], $v_{mic}$, $v_{mac}$, as well as abundances of C, N, Mg, Si, and Ti were all kept as free parameters in PySME while fitting parts of the spectra covering temperature-sensitive OH, CO, and CN molecular lines and band heads, and atomic lines of C I, Mg I, Si I, Ti I, Ti II, and Fe I. Regarding $\log g$, it was not a free parameter in the spectrum fitting; rather, it is a so called "derived parameter" in PySME, and is calculated based on the age, mass, distance, visual extinction ($A_V$), and G magnitude using photometric approaches while the other parameters mentioned above are optimized. The method combines the theoretical isochrones with observational data to derive $\log g$. The selection of an appropriate isochrone is based on the age and [Fe/H] of the star in question, with the MIST model grid (Paxton et al. 2015; Dotter 2016; Choi et al. 2016) being utilised for this purpose. The point on the isochrone that corresponds to the fitted effective temperature is identified to get a mass-estimate for the star. The absolute G magnitude is computed by utilising the observed G-band magnitude of the star, the distance to the cluster, and the cluster reddening (Cantat-Gaudin et al. 2020; Brogaard et al. 2021). The interstellar reddening was calculated by converting the visual extinction $A_V$ to color excess $E(B - V)$ using the standard total-to-selective extinction ratio $R_V = 3.1$, following the extinction law of Cardelli et al. (1989). A bolometric correction is then applied (Casagrande & VandenBerg 2018a,b), to ultimately yield the bolometric magnitude and, consequently, the stellar luminosity. The surface gravity is then finally calculated using the following relation:

$$\log g = \log g_\odot + \log\left(\frac{M}{M_\odot}\right) - \log\left(\frac{L}{L_\odot}\right) + 4\log\left(\frac{T_{eff}}{T_\odot}\right) \quad (1)$$

where solar reference values are $\log g_\odot = 4.44$, $M = 1 M_\odot$, and $T_\odot = 5777$ K.

Given that the present stellar sample includes both red giant branch and red clump stars, this method provides a consistent and reliable approach for estimating $\log g$ across different evolutionary phases (as opposed to the isochrone-method used in for example Bijavara Seshashayana et al. 2024b).

Similarly, [O/Fe] is set to be a PySME "derived parameter", following [Mg/Fe]. This approach is motivated by the shared nucleosynthetic origin of (the lightest) $\alpha$ elements, which are almost entirely produced in core-collapse supernovae. In stars that are too warm to exhibit any OH-lines, direct measurements for constraining O are not possible. Therefore, aligning [O/Fe] with [Mg/Fe] helps maintain a physically motivated chemical pattern (Woosley & Weaver 1995; Nomoto et al. 2013; Kobayashi et al. 2020).

The stellar parameters derived for all 114 stars from 41 clusters are shown as Kiel diagrams in Figures C.1-C.2. These Figures presents our sample data alongside stellar parameters from Dal Ponte et al. (2025), including both overlapping stars and those unique to each dataset. The corresponding isochrones of the same age and metallicity as the clusters are taken from the MIST database.

The atomic and molecular lines utilized to derive the stellar parameters are presented in Table B.1, while the derived stellar parameters can be found in Table C.1. This table includes the formal fit uncertainties from PySME for each parameter across all stars. The average uncertainties for $T_{eff}$, $\log g$, [Fe/H], $v_{mic}$, and $v_{mac}$ are 18 K, 0.1 dex, 0.02 dex, 0.07 km/s, and 0.17 km/s, respectively. These uncertainties reflect the random variation associated with the stellar parameters, arising from factors such as S/N, potential issues with $\chi^2$ minimization, and to some extent, continuum normalization. In contrast, the often significantly larger systematic uncertainties, which originate from aspects like model atmospheres and atomic data, remain much more challenging to quantify. Since $\log g$ was determined using isochrones, the uncertainties were estimated by propagating errors in $T_{eff}$ (fit uncertainties) through the standard relation for $\log g$ based on stellar parameters. A comparison of our determined stellar parameters with findings from a few other studies is discussed in Section 4.2.

### 3.2. Elemental abundances

The next step in the analysis is to measure the elemental abundances, adopting the determined stellar parameters. This is also done using PySME to vary the abundance in question to optimize the fit of the synthetic spectrum to the observed. One element at the time was fitted, and in total up to 23 elements were determined in every star.

For some spectra, the analysis was not applied to specific elements due to the unavailability of measurable lines, too low S/N ratio, or significant telluric contamination. The number of stars included in the determination of each element are listed in Table D.1.

The values of [X/Fe] for all analyzed elements in the individual stars are presented in Tables D.2-D.3, along with their formal fit uncertainties from PySME. Similar to the stellar parameters, these uncertainties reflect random variations primarily caused by factors such as the S/N ratio. Systematic uncertainties in stellar abundances typically surpass these random errors and are mainly driven by inaccuracies in stellar parameters and atomic data. To mitigate the influence of atomic data inaccuracies, we have utilized rigorously vetted and astrophysically adjusted atomic data, as detailed above. The mean cluster abundances listed in Tables D.4-D.6 demonstrate intra-cluster variations generally ranging from 0.01 to 0.10 dex, with only a few cases reaching 0.15 dex, thereby reinforcing the reliability of our abundance determinations.

All our abundance results are compared with [Fe/H], age, and $R_{gc}$ in Figures 1–6. These trends, visible for individual elements, provide valuable insights into the chemical evolution and spatial structure of the Milky Way's disc population.

## 4. Results and Discussion

The derived abundances for each cluster is listed in Tables D.4-D.6, and the abundance trends are plotted compared with [Fe/H], age, and $R_{gc}$ in Figures 1–6. The quoted uncertainties represent the error on the mean abundance $\left(\sigma/\sqrt{N}\right)$, except for clusters with only one observed star. In such cases, representative uncertainties were estimated based on the average uncertainties from clusters with multiple stars. All abundance ratios are scaled relative to the solar values as reported by Asplund et al. (2021).

The [X/H] versus $R_{gc}$ slopes for all elements in our sample and many of the literature values discussed below are listed in Table 1.

### 4.1. A special cluster: NGC 6791

NGC 6791 is considered to be among the oldest OC of the Galaxy. In their seminal work, Brogaard et al. (2021), definitively established an age of 8.3±0.3 Gyr, a conclusion that was reached through the meticulous analysis of multiple eclipsing binaries. This is the most robust determination and should be preferred to the values in Cantat-Gaudin et al. (2020) or Hunt & Reffert (2023). The metallicity of this cluster has been determined by multiple authors using both photometry (e.g., Brogaard et al. 2021, who found a metallicity of 0.3), and spectroscopy, with values ranging from approximately 0.2 to over 0.4 dex (e.g., Casamiquela et al. 2017; Gratton et al. 2006, respectively). Our [Fe/H] value of 0.28±0.03 dex is in very good agreement with previous values, including Bijavara Seshashayana et al. (2024a).

Given its very high metallicity and its current $R_{gc}$, it has long been suspected that NGC 6791 formed much closer to the Galactic centre before migrating outward. For example, Jílková et al. (2012) proposed a migration from an inner-disk position ($R_{gc}$ of 3–5 kpc) to the current value of approximately 8 kpc. Additionally, its large orbital eccentricity and distance from the Galactic plane, coupled with its metallicity and [$\alpha$/Fe] greater than solar (see, for example, Viscasillas Vázquez et al. 2022), make this cluster an outlier. Hence, NGC 6791 has been excluded from all slope calculations presented in the following subsections and figures, in order to avoid skewing the derived abundance trends.

### 4.2. Comparison of parameters and abundances with literature values

The majority of the OCs analysed in this study have already been investigated in previous works based on optical spectra from the SPA project, including Casali et al. (2020), Zhang et al. (2021), Zhang et al. (2022), Alonso-Santiago et al. (2021), and Dal Ponte et al. (2025). There are also OCs from our sample in common with studies, such as OC-CASO (Casamiquela et al. 2017, 2019), see below. Furthermore, the OCCAM-project within the APOGEE survey has examined a sample of stars in 150 clusters in the IR region (Myers et al. 2022), with nine of these clusters overlapping with our sample.

In order to evaluate the consistency and discrepancies between our analysis and existing literature, we compared our stellar parameters, $T_{eff}$, $\log g$, and [Fe/H] — for stars in common with other datasets. The differences were calculated on a star-by-star basis as the mean difference (this work - literature) per cluster, with uncertainties expressed as standard deviations. As demonstrated in Table C.2, this reveals a range of discrepancies between this work and previous studies. The offsets are generally modest (typically $\leq$ 100 K in $T_{eff}$, 0.3 dex in $\log g$, and 0.1 dex in [Fe/H]). However, the differences in $\log g$ are sometimes larger, reflecting the general difficulty of determining this parameter accurately. In a number of cases, and especially in comparison with Zhang et al. (2021), the discrepancies are even more marked, with one star displaying a difference of 1.56 dex in $\log g$. These larger offsets likely arise from the differing methods used to derive $\log g$: Zhang et al. (2021) determined $\log g$ spectroscopically from ionization balance, whereas in the present work, $\log g$ was computed photometrically by combining isochrone-based mass estimates with cluster distances and extinction corrections. Additional differences in adopted line lists, model atmospheres, and reference solar abundances may also contribute to the systematic offsets. When it comes to our SPA-companion-paper Dal Ponte et al. (2025), the derived stellar parameters are however in general very similar. Since our analysis is based on IR data, whereas Dal Ponte et al. (2025) utilised optical spectra this strong agreement underscores the reliability of the derived stellar parameters.

To contextualize our IR elemental abundance results, we compare our data primarily with the OCCAM project within APOGEE (DR17, Myers et al. 2022), which is based on high-resolution H-band spectra analyzed using an industrial pipeline (ASPCAP) and is thus directly comparable to our work in terms of wavelength coverage, stellar parameter sensitivity, and potential systematics. Since our analysis is based on IR spectra, we deliberately avoid combining our results with optical-only surveys unless explicitly stated. However, we include Dal Ponte et al. (2025) in our plots as a relevant optical counterpart, as being a SPA paper, also based on observations from the TNG and target similar stellar populations. In specific cases where IR data are lacking for certain elements or clusters, we include results in the discussion section only and not in plots, from well-characterized optical surveys such as OCCASO (Carbajo-Hijarrubia et al. 2024) and Gaia-ESO (Magrini et al. 2023).

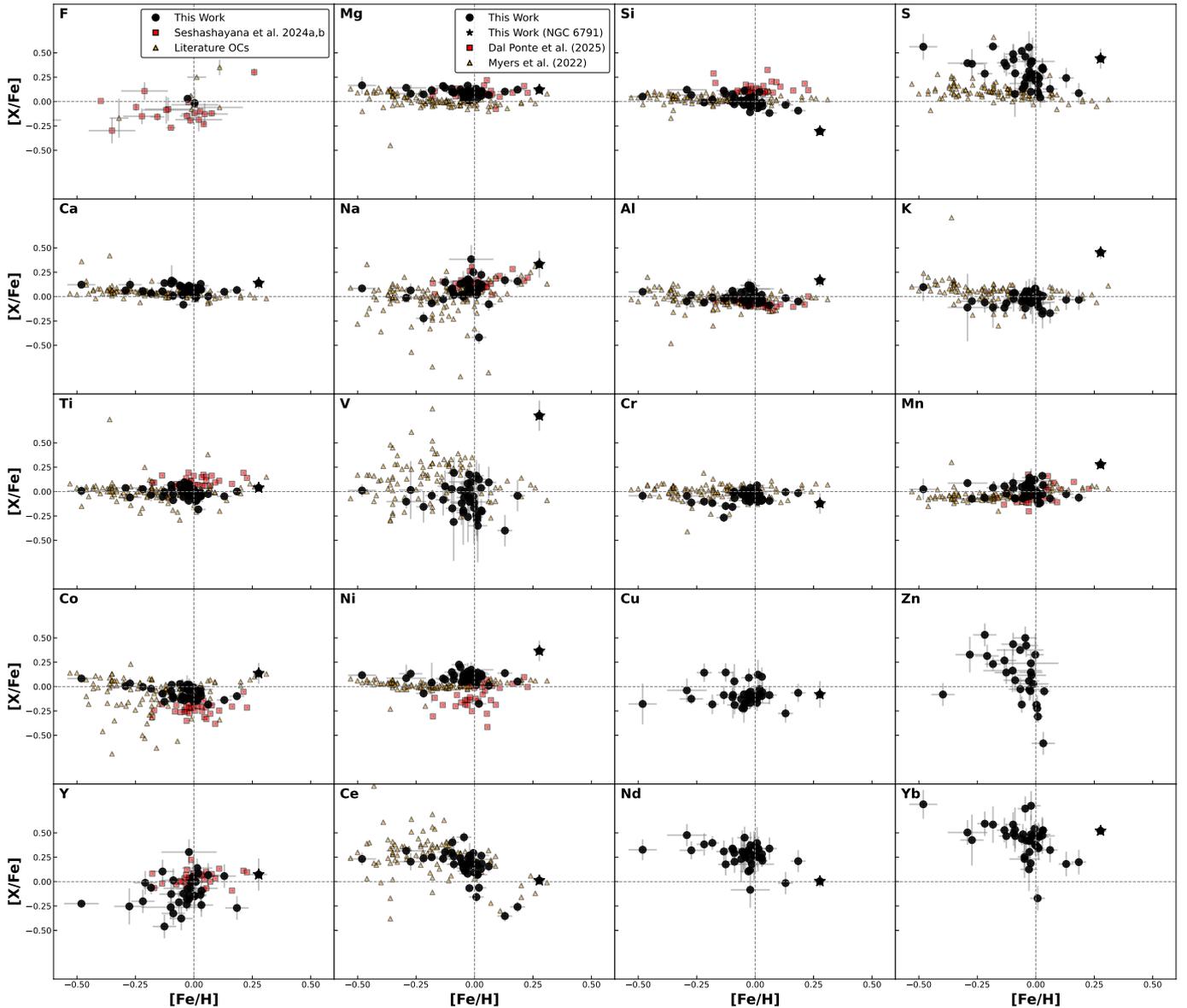

Fig. 1: Mean cluster abundance ratios as a function of [Fe/H]. The black markers are from the present work, with the black star indicating the open cluster NGC 6791. The red squares are from Dal Ponte et al. (2025), and orange triangles are from Myers et al. (2022) for all elements but F. For F, red squares are from Bijavara Seshashayana et al. (2024a); Bijavara Seshashayana et al. (2024b), orange triangles are all other literature OCs with determined F abundances (for more information, see Bijavara Seshashayana et al. 2024a; Bijavara Seshashayana et al. 2024b).

Cross-validating results across these wavelength regimes enhances confidence in the derived chemical signatures and their application to tracing the chemical enrichment of the Milky Way.

For the [X/Fe] values, we compared with Dal Ponte et al. (2025) and APOGEE DR17 (Myers et al. 2022). The comparison with Dal Ponte et al. (2025) was carried out using 29 clusters and for Mg, Si, Ti, Na, Al, Mn, Co, Ni, and Y. There is a general consensus for the majority of species. The median absolute difference ($|\Delta[X/Fe]|$) is typically $\leq$0.07 dex for $\alpha$-elements and for Fe-peak elements, suggesting a consistent abundance scale despite methodological differences. Larger, though still systematic, offsets were observed for the neutron-capture element Y. This may be attributed to the use of NLTE corrections in Dal Ponte et al. (2025), as opposed to the LTE assumptions adopted in this work. The element Na displays only negligible star-to-star scatter, suggesting that its abundance remains relatively consistent across the sample (approximately 0.08–0.10 dex).

In the case of APOGEE DR17, 14 elements overlap with our work: Na, Mg, Al, Si, S, K, Ca, Ti, V, Cr, Mn, Co, Ni, and Ce. The agreement is encouraging, with the majority of $\Delta[X/Fe]$ values falling within $\pm$0.05 dex.

Despite inherent differences in wavelength coverage, data reduction techniques, and analysis pipelines, the abundance patterns derived from the three independent datasets

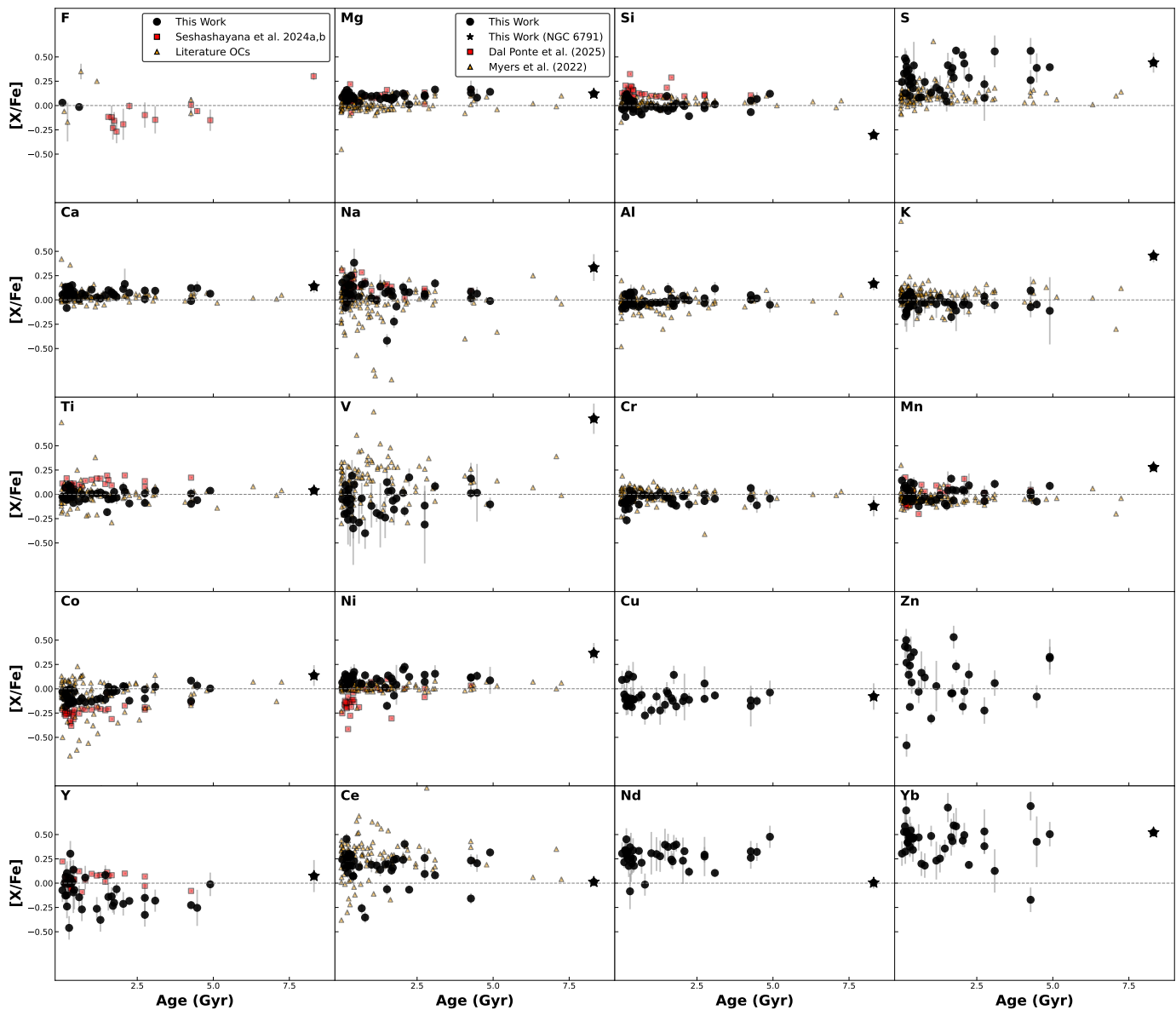

Fig. 2: Same as Figure 1 but here [X/Fe] values are plotted against age.

– ours, Myers et al. (2022), and Dal Ponte et al. (2025) – exhibit broad concordance. For clusters common to our study and Dal Ponte et al. (2025) – such as Alessi 1, Basel 11b, and Gulliver 18 – the mean abundance differences in key elements including Mg, Si, Ca, and Fe typically lie within ±0.05 dex. It is notable that these offsets are comparable to, or smaller than, the combined uncertainties of the measurements. Some minor systematic deviations are apparent, such as slightly lower Mg and Si abundances in our data for Alessi Teutsch 11 relative to Dal Ponte et al. (2025), and modest differences in S and Na for select clusters. Although our analysis yields very small formal uncertainties, these mainly reflect the fitting procedure and do not fully account for systematic effects.

A more representative measure of precision is the internal abundance scatter within a cluster, which we use for comparison across studies. The close correspondence in abundance trends between the two datasets provides com-pelling evidence for the robustness of the analytical approach. The minor residual discrepancies can be attributed to variations in line lists, analytical methods, or the criteria employed for sample selection.

The mean cluster abundances obtained in this study also demonstrate excellent agreement with those reported by Myers et al. (2022), with mean differences across key elements typically within ±0.02 dex. However, it is important to note that the internal uncertainties representing cluster abundance spreads in our analysis are on average approximately 40% smaller than those presented by Myers et al. (2022). This substantial reduction in internal dispersion is consistent across all elements studied and attests to the enhanced precision and robustness of the methodology employed.

Our derived radial metallicity gradient is −0.079 ± 0.018 dex/kpc including NGC 6791 and −0.072 ± 0.017 dex/kpc excluding NGC 6791 (see Figure 7), aligning well with val-ues reported in recent OC studies. For instance, Myers et al. (2022) found a similar slope of $-0.073\pm0.002$ dex/kpc within the range of 6–11.5 kpc and Spina et al. (2021) reported $-0.076 \pm 0.009$ dex/kpc and these clusters are located at R$_{gc}$ spanning approximately 6 to 16 kpc. These consistent findings across various datasets and methodologies reinforce the existence of a negative abundance gradient in the inner Galactic disc. The decrease of metallicity with R$_{gc}$ is a well-established result, consistently confirmed over decades of OC studies. Beyond the distance of 11–13 kpc, a noticeable flattening of the gradient is frequently observed, with slopes tending toward zero Magrini et al. (2023). The findings of this study emphasise the robustness of the metallicity gradient and lend support to models of inside-out disc formation shaped by radial migration and external accretion. Recio-Blanco et al. (2023) utilised the Gaia-RVS to derive the metallicity gradient for almost 500 OCs, thereby further reinforcing the robustness of this re-sult. Whilst it is acknowledged that only a limited number of studies have been referenced in this analysis, the general consensus is one of consistency. Another valuable approach to probing the properties of the Galactic disc using OCs is presented by Hu & Soubiran (2025), who utilised Gaia spectrophotometry to analyse a sample of approximately 600 OCs covering a wide range of distances and ages. A precise radial metallicity gradient was derived, which revealed a break at R$_{gc}$ 10-12 kpc: the inner disc exhibits a steep slope of $-0.084\pm0.004$ dex/kpc, while the outer disc shows a significantly flatter gradient of $-0.018\pm0.056$ dex/kpc. In addition, they reported a vertical gradient of $-0.415\pm0.030$ dex/kpc, highlighting the disc's chemical stratification with height above the Galactic plane. We note that only a small subset of the extensive literature on metallicity gradients is presented here.

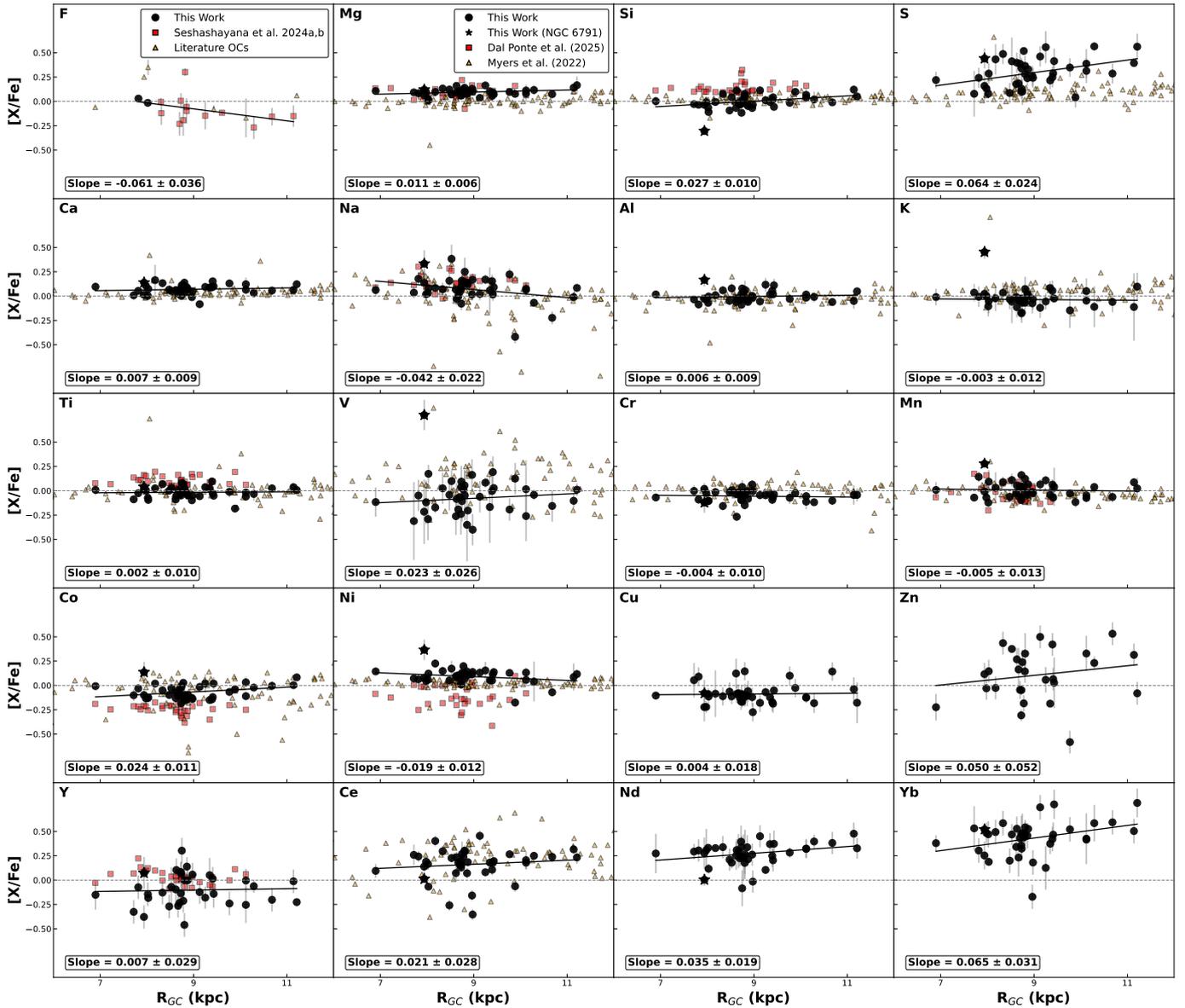

Fig. 3: Same as Figure 1 but here [X/Fe] values are plotted against R$_{gc}$.

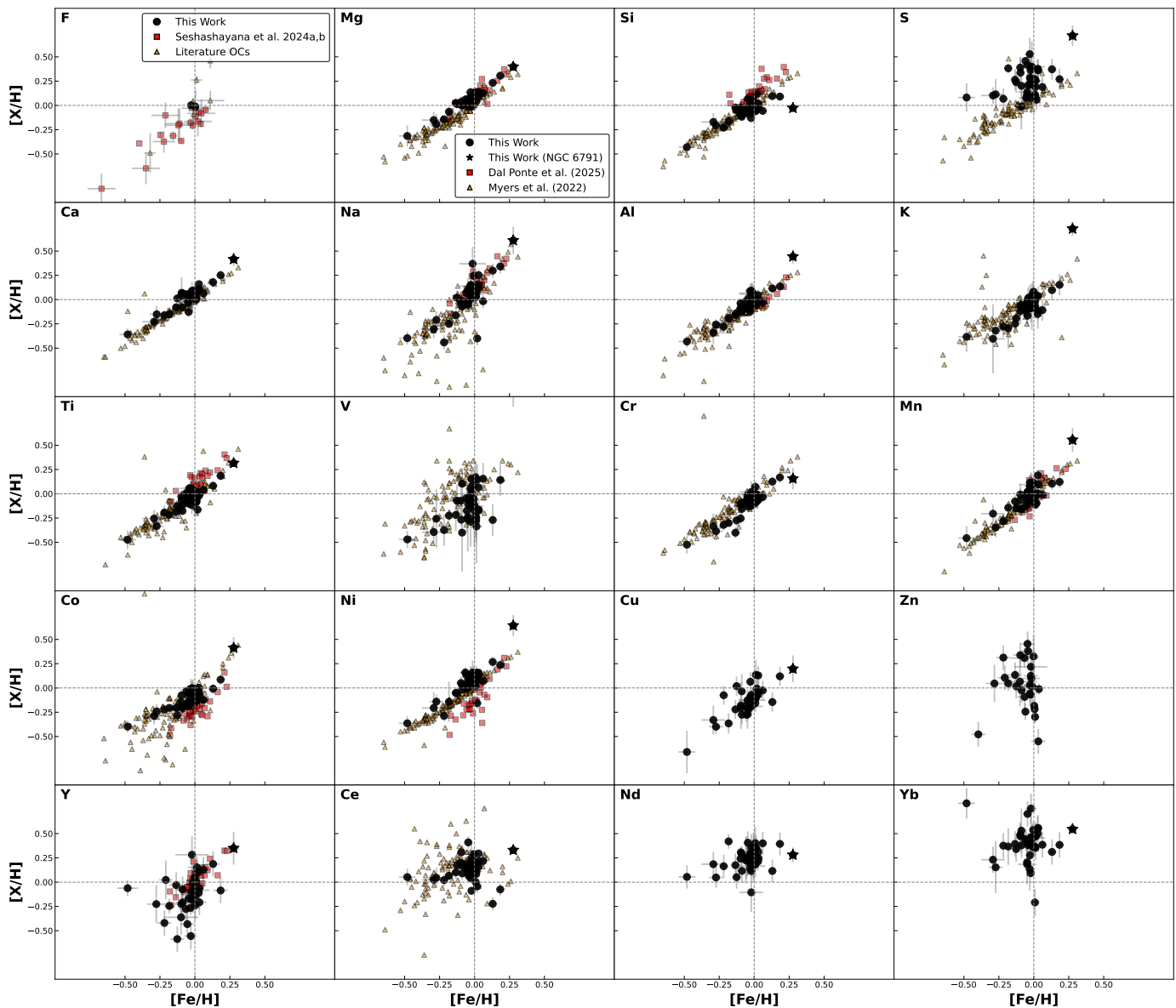

Fig. 4: Abundance [X/H] ratios as a function of [Fe/H]. Symbols are as in Figure 1.

### 4.3. α elements (Mg, Si, S, Ca, and Ti)

The α elements are among the most studied elements in the context of Galactic chemical evolution. They are mainly synthesised in core-collapse supernovae (CCSNe), which originate from massive stars with masses $\gtrsim 10\ M_\odot$. Due to their short lifetimes, these stars rapidly enrich the interstellar medium (ISM) with α elements early in the history of the Galaxy. In contrast, iron is produced predominantly in Type Ia supernovae, which are produced by lower-mass progenitors and occur on longer timescales. This temporal difference leads to the well-known enrichment of [α/Fe] at low metallicities (Tinsley 1980; Matteucci & Greggio 1986).

Magnesium: The Mg I abundances were determined using four spectral lines in the H-band (for details, see Table B.1). NLTE synthesis was applied to all the Mg lines used, but a LTE synthesis would only render 0.03 dex higher [Mg/Fe] on average. Mg abundances were derived for all clusters in our sample, and the results are presented in Figures 1-6, where they are plotted and compared with literature values from Dal Ponte et al. (2025) and Myers et al. (2022). The [Mg/Fe] ratio shows a flat trend when plotted against $R_{gc}$ (slope: $0.011 \pm 0.006$ dex/kpc), as expected for an α-element like Mg, which is produced predominantly in CCSNe early in the Galaxy's history, leading to broadly uniform enrichment across the disc. The trend of [Mg/H] with $R_{gc}$ reveals a clear negative gradient ($-0.061 \pm 0.016$ dex/kpc), consistent with the metallicity gradient expected from inside-out disc formation. Our [Mg/Fe] measurement agrees with the Gaia–ESO slope of $0.008 \pm 0.006$ dex/kpc (Magrini et al. 2023) and with the OCCASO slope of $0.017 \pm 0.011$ dex/kpc (Carbajo-Hijarrubia et al. 2024). The overall agreement between our results and these previous studies supports the view that Mg, as a typical α-element, is well-mixed throughout the

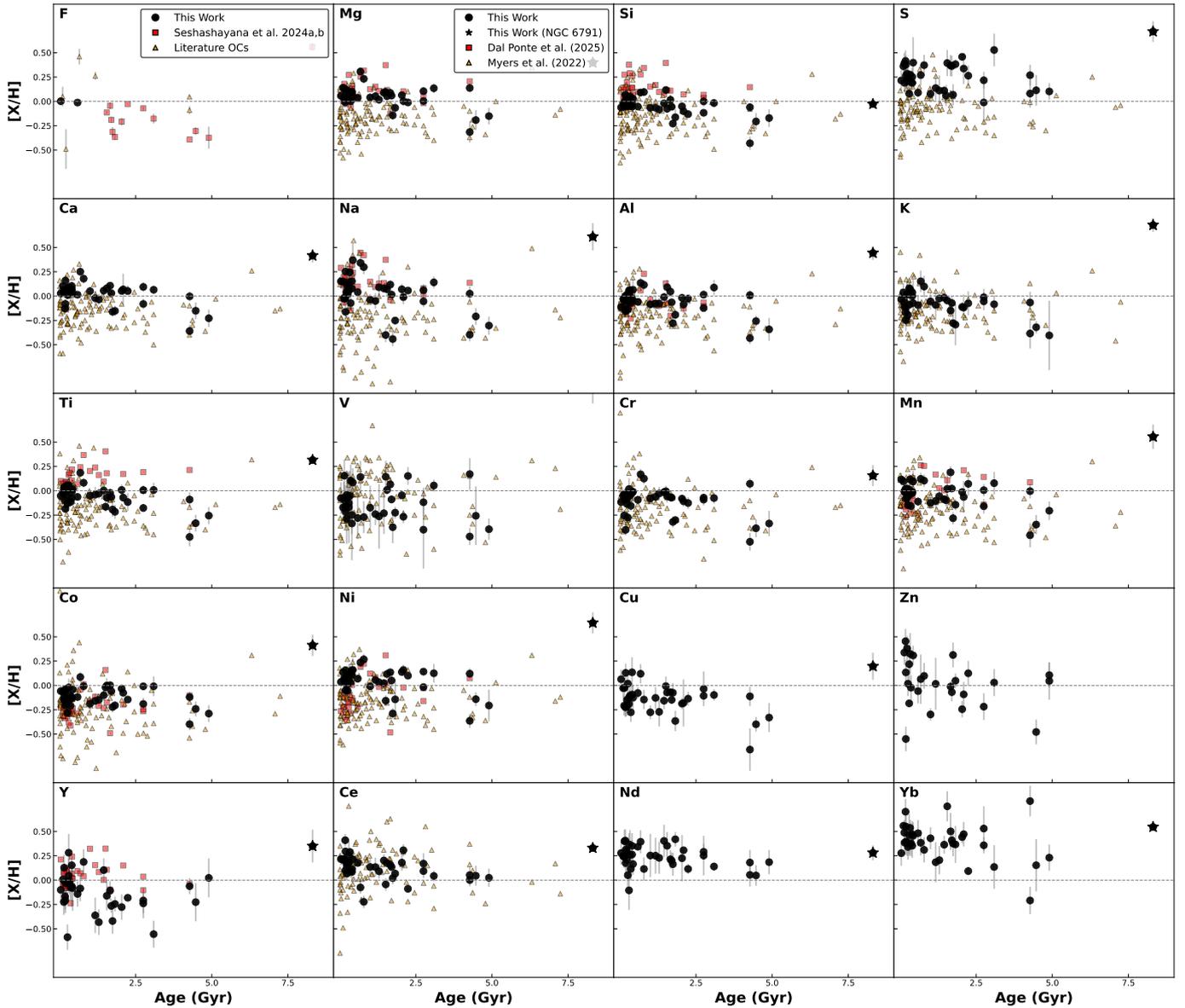

Fig. 5: Same as Figure 4 but here [X/H] values are plotted against age.

Galactic disc and does not display strong spatial variation.

Silicon: The Si I abundances were determined using nine spectral lines in the H-band region (see Table B.1). NLTE synthesis was applied to all the Si lines used, but a LTE synthesis would render 0.05 dex higher [Si/Fe] on average. We plot and compare our Si abundance results with those of Myers et al. (2022) and Dal Ponte et al. (2025), finding good agreement across the datasets. [Si/Fe] remains largely flat across [Fe/H] and age, with only a slight positive trend. This suggests that Si production, like Mg, is closely tied to that of Fe over the evolution of the disc, though it may exhibit slight enhancements in the outer regions compared to the inner disc. [Si/H] displays a negative slope with R$_{gc}$ ($-0.044 \pm 0.016$ dex/kpc), again tracing the underlying metallicity gradient. The comparatively shallower gradient of [Si/H] with R$_{gc}$ relative to that of [Mg/H] may reflect fundamental differences in their nucleosynthetic origins. While both elements are primarily produced in CCSNe, Mg is synthesised predominantly during hydrostatic burning in massive stars, whereas Si is formed largely through explosive O burning, making its yield more sensitive to progenitor mass and explosion energy (e.g. Woosley & Weaver 1995; Nomoto et al. 2013). Furthermore, Si has a modest contribution from Type Ia supernovae (e.g. Iwamoto et al. 1999; Kobayashi et al. 2020), which produce both Fe and Si but negligible Mg. This additional Si contribution acts to flatten the [Si/Fe] gradient, especially in the inner disc where the SNe Ia rate is higher. Conversely, the outer disc, where the occurrence of SNe Ia is less prevalent, exhibits a more pronounced retention of pristine $\alpha$-element signatures. However, the [Si/Fe] gradient in this region remains less pronounced in comparison to that of [Mg/Fe]. For [Si/Fe] versus R$_{gc}$, our data yield a slope of $0.027 \pm 0.010$ dex/kpc, which is more positive than most previous studies. For instance, Myers et al. (2022) report values clustered around $0.011 \pm 0.001$ dex/kpc. The value reported by Magrini et al. (2023) in Gaia-ESO is $0.003 \pm 0.010$. Carbajo-Hijarrubia et al. (2024) reported a value of $0.003 \pm 0.003$. The more positive trend observed in our sample could suggest a mild enhancement of Si in the outer disc. This may reflect variations in the star formation history, the initial mass function, or the role of Type II supernovae at larger radii. However, given the error bars, the difference between our result and those in the literature is not significant, and the overall [Si/Fe] gradient remains weak. This supports the broad conclusion of a chemically homogeneous thin disc for Si.

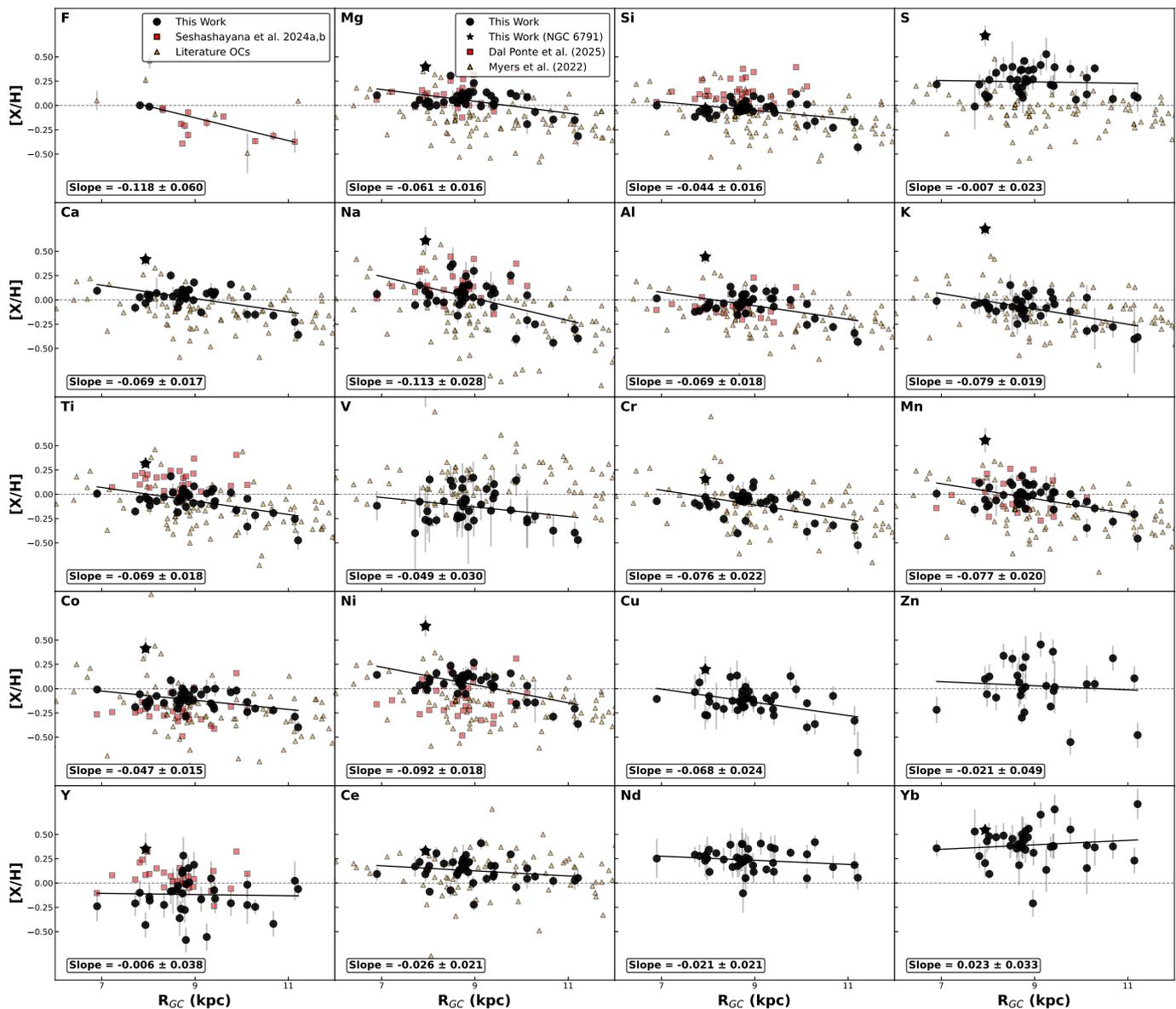

Fig. 6: Same as the figure 4 but here [X/H] values are plotted against R$_{gc}$.

Sulphur: We derived S abundances using only one S I line at 15478.48 Å. NLTE synthesis was applied to the S line, and a LTE synthesis would render 0.06 dex lower [S/Fe] on average, indicating non-negligible NLTE-effects in our sample of stars. There are relatively few studies on S abundances in stars, partly due to the challenges associated with these NLTE sensitivities. In our sample, we rely on a single S line, which introduces larger uncertainties in the abundance measurements. Despite this, when uncertainties are taken into account, the [S/Fe] ratio shows a positive trend with R$_{gc}$ ($0.064 \pm 0.024$ dex/kpc), implying a possible slight enhancement in the outer disc. Our S abundances are systematically larger with those reported by Myers et al. (2022) as seen in Figures 1-6. The [S/H] trend with R$_{gc}$ is nearly flat, with a weak negative slope ($-0.007 \pm 0.023$ dex/kpc), suggesting that the absolute abundance decreases only slightly with increasing R$_{gc}$. This behaviour is somewhat unexpected, given that S is traditionally considered an $\alpha$-element and should therefore trace the trends seen for elements like Mg or Si. The nearly flat [S/H] gradient exhibited in this study stands in contrast to the results obtained from Cepheid studies, including those by da Silva et al. (2023) and Trentin et al.

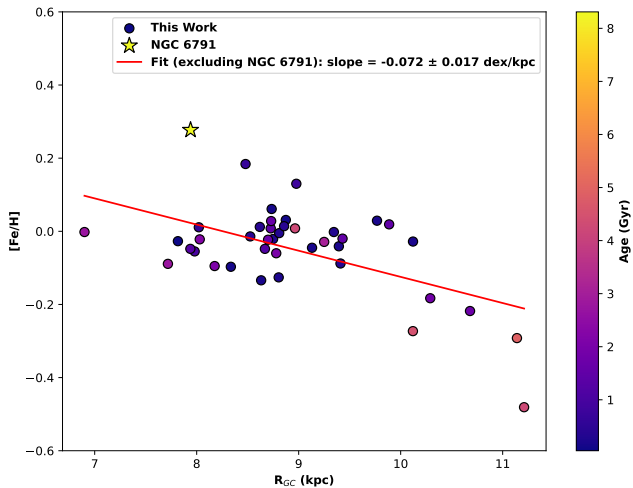

Fig. 7: [Fe/H] vs. R$_{gc}$ for our clusters. Each point is colour-coded according to cluster age (in Gyr), as indicated by the colour bar. The red line indicates the linear fit, with the exclusion of NGC 6791 (denoted by a star symbol). The derived slope and its uncertainty are indicated in the legend.

(2024). These studies reported significantly steeper positive [S/H] gradients with R$_{gc}$ for young stellar populations. One plausible explanation for this discrepancy lies in the age difference between the stellar populations being probed. Cepheid variables are relatively young, massive stars, with typical ages of only a few hundred million years. The OCs in our sample, on the other hand, span intermediate to old ages, thereby tracing earlier phases of the Galactic disk evolution. While Cepheids provide information on the most recent chemical enrichment, OCs offer a complementary, longer-term perspective. A steeper [S/H] gradient in the younger Cepheid population would suggest that S production (primarily from CCSNe) has become more concentrated towards the center in recent times. In other words, the outer disc has received relatively less S enrichment. Alternatively, metallicity-dependent yields and diminished radial gas mixing could have steepened the gradient over time (e.g. Kobayashi et al. 2006; Prantzos et al. 2018). Conversely, it can be hypothesised that the results obtained from our OC sample may be indicative of an earlier phase of Galactic chemical evolution during which S was present but distributed differently across the disc (e.g. Matteucci 2001; Romano et al. 2010).

Calcium: Ca abundances were determined using seven H- and K-band lines (see Table B.1). NLTE synthesis was applied to all the Ca lines used, but a LTE synthesis would only render 0.01 dex higher [Ca/Fe] on average. One star, UBC_141_1, was excluded from the analysis, due to a low S/N ratio that rendered the Ca-lines unusable. We plot and compare our Ca abundance results with those from Myers et al. (2022) and Dal Ponte et al. (2025), which show good agreement. [Ca/Fe] shows no significant trend with R$_{gc}$ (0.007 ± 0.009 dex/kpc), reflecting Ca being an $\alpha$-element produced in similar sites as Mg and Si. This finding is fully consistent with previous literature [Ca/Fe] values, including the studies by Reddy et al. (2016) (0.014 ± 0.005 dex/kpc), Magrini et al. (2023)

(0.010 ± 0.007 dex/kpc), and Carbajo-Hijarrubia et al. (2024) (0.015 ± 0.005 dex/kpc) all of which reported very weak or negligible gradients for [$\alpha$/Fe] versus R$_{gc}$. [Ca/H], however, decreases with increasing R$_{gc}$ (−0.069 ± 0.017 dex/kpc), consistent with the overall negative metallicity gradient of the Galactic disc. The near-zero slope of [Ca/Fe] supports the idea that the level of enrichment relative to iron has remained consistent across the disc among the $\alpha$-elements, reflecting a well-mixed interstellar medium and comparable supernovae yields in various disc environments.

Titanium: Ti abundances were determined using five IR lines as shown in the table B.1 using both Ti I and Ti II lines. All stars in the sample were included in the analysis, and NLTE synthesis was applied to all the Ti lines used, but a LTE synthesis would only render 0.01 dex higher [Ti/Fe] on average. Our results are plotted with those from Dal Ponte et al. (2025) and Myers et al. (2022), showing consistent trends. The [Ti/Fe] ratios display a flat behavior as a function of R$_{gc}$, suggesting a steady production ratio over time and across different Galactic regions. This result is fully consistent with the findings of Myers et al. (2022), Magrini et al. (2023), and Carbajo-Hijarrubia et al. (2024) who report negligible gradients for [Ti/Fe] across the disc. [Ti/H] decreases with R$_{gc}$ (−0.069 ± 0.018 dex/kpc), matching the metallicity gradient. These trends reflect the expected behavior of Ti as an element with contributions from CCSNe, becoming more enriched in regions closer to the Galactic center. The lack of a significant [Ti/Fe] gradient suggests that Ti, like many other $\alpha$-elements, is well-mixed in the Galactic thin disc, and its enrichment relative to Fe is uniform over large spatial scales.

In summary, the findings of this study demonstrate that the measurements for [Mg/Fe], [Si/Fe], [S/Fe], [Ca/Fe], and [Ti/Fe] versus R$_{gc}$ are consistent with the prevailing consensus in the existing literature. The $\alpha$-element abundances relative to iron remain relatively constant across the disc, exhibiting at most mild positive gradients for Si and S, and virtually no gradient for Mg and Ca. Whilst the overall trends are broadly similar to ours, the [X/Fe] values from Myers et al. (2022) exhibit significantly greater scatter across [Fe/H], age, and R$_{gc}$ as seen from Figures 1-6. This is likely attributable to the industrial analysis approach in OCCAM and the absence of uniform NLTE corrections, both of which introduce uncertainties and inflate the observed dispersion. In contrast, the measurements presented in this study show tighter clustering and more well-defined trends. This reflects a more homogeneous analysis based on high-S/N spectra and carefully selected spectral lines. Notably, our results are in strong agreement with Dal Ponte et al. (2025) – the optical counterpart of this study – even though different methodologies and line data were employed, underscoring the robustness of the derived abundance patterns. These findings support the hypothesis of a chemically homogeneous thin disc for the elements considered here.

### 4.4. Odd-Z elements (F, Na, Al, and K)

We analyzed F, Na, Al, and K as representatives of the odd-Z elements. These elements are believed to be produced in both massive stars and asymptotic giant branch (AGB) stars. Their abundance can exhibit secondary behaviour, meaning their production depends on the initial metallicity of the stellar environment. As a result, such abundances tend to increase with metallicity, showing higher values at higher [Fe/H] (Smiljanic et al. 2016).

Fluorine: Continuing from two previous GIANO-B-based studies (Bijavara Seshashayana et al. 2024a; Bijavara Seshashayana et al. (2024b)), F abundances were determined for two additional OCs in the present work: Collinder 350 and Gulliver 18. The analysis was performed using the HF R9 molecular line at 23358.33 Å. Although widely adopted for F abundance studies, this line can suffer from blending and saturation in cool ($T_{eff}$ < 3500 K) and metal-rich ([Fe/H] > 0.0 dex) stars (see Nandakumar et al. 2023b); however, this does not affect our present analysis, as the stars analyzed in this paper are warmer ($T_{eff}$ > 4000 K). Our derived F abundances are in good agreement with both the literature sample of OCs and those analyzed in our earlier work. For further details on the methodology and comparative Galactic trends, we refer to Bijavara Seshashayana et al. (2024a); Bijavara Seshashayana et al. (2024b).

Sodium: Na abundances were derived using a combination of H- and K-band spectral lines (Table B.1). NLTE synthesis was applied to the Na lines used, but a LTE synthesis would render 0.20 dex higher [Na/Fe] on average, indicating significant NLTE-effects for our Na-lines in our sample of stars. Three stars – NGC_2509_18, NGC_2420, and Gulliver_24 – stand out with slightly lower Na abundances and larger associated uncertainties, likely due to spectral quality. The [Na/Fe] shows a mild negative trend with $R_{gc}$ ($-0.042 \pm 0.022$ dex/kpc), indicating that Na enrichment relative to iron may decrease with distance from the Galactic center. This trend is in decent agreement with previous studies for [Na/Fe] values, such as Myers et al. (2022), with $-0.006 \pm 0.007$ dex/kpc, Magrini et al. (2023), with $0.003 \pm 0.002$ dex/kpc, and Carbajo-Hijarrubia et al. (2024) with $-0.027 \pm 0.008$ dex/kpc. All of these literature values are consistent with our result within the stated uncertainties. The [Na/H] trend with $R_{gc}$ is stronger, with a slope of $-0.113 \pm 0.028$ dex/kpc, indicating a strong decline outward in the disc, reinforcing the trend of increasing Na enrichment toward the inner Galaxy. The presence of a negative gradient in the [Na/Fe] ratio with respect to $R_{gc}$ may suggest that nucleosynthetic yields are metallicity-dependent. We also note a marginally higher [Na/Fe] ratio in younger OCs, consistent with previous findings (e.g. Smiljanic et al. 2016) that attribute such trends to evolutionary effects in giant stars. This behaviour persists after excluding NGC 6791, which is chemically peculiar and significantly older than the rest of the sample. The observed trend could reflect variations in the evolution of massive stars or differences in Type II supernova contributions in the outer disc. However, this interpretation is complicated by the age–$R_{gc}$ degeneracy, as older clusters are preferentially located at larger $R_{gc}$ in our sample. This spatial-age correlation highlights the challenge of disentangling evolutionary effects from dynamical influences when interpreting radial abundance gradients.

Aluminium: Al abundances were determined using six spectral lines located in both H- and K-band regions.

NLTE synthesis was applied to all the Al lines used, but a LTE synthesis would render 0.09 dex lower [Al/Fe] on average, indicating modest NLTE-effects for our Na-lines in our sample of stars. [Al/Fe] vs. $R_{gc}$ does not display a significant trend ($0.006 \pm 0.009$ dex/kpc). Our results are plotted with those from Dal Ponte et al. (2025) and Myers et al. (2022), showing good agreement across the sample. Carbajo-Hijarrubia et al. (2024) reports a value of $-0.001 \pm 0.005$ dex/kpc which is also in extremely good agreement with us. The [Al/Fe] value reported by the Gaia-ESO Survey (Magrini et al. 2023), $0.012 \pm 0.004$ dex/kpc, is in agreement with our result within the uncertainties. [Al/H], by contrast, shows a marked negative slope with $R_{gc}$ ($-0.069 \pm 0.018$ dex/kpc), similar to other metallicity indicators. This finding is consistent with most recent literature, which also reports flat gradients for [Al/Fe] across the Galactic disc. The lack of a significant [Al/Fe] gradient suggests that Al enrichment relative to Fe has remained uniform with increasing distance from the Galactic center, possibly reflecting well-mixed nucleosynthetic contributions from both CCSNe and Type Ia supernovae throughout the disc's history.

Potassium: K abundances were derived using one spectral line at 15168.40 Å. All clusters in our sample were analyzed for K, and NLTE synthesis was applied to all the K line, but a LTE synthesis would render 0.03 dex lower [K/Fe] on average. K has not been extensively studied in stellar abundance analyses, mainly due to observational challenges. In the optical, the strongest K I lines (at 7665 Å and 7699 Å) are often avoided because they are affected by interstellar absorption, lie in regions contaminated by telluric lines, and are typically strong and saturated, making them difficult to model accurately. In the IR, the 15168.40 Å line is a viable alternative, but it is one of the few usable K lines and is often weak or mixed, especially in cool or metal-poor stars. Our results are primarily plotted with those from Myers et al. (2022), showing good agreement. The Galactic trends observed in our sample align well with expectations from chemical evolution models. [K/Fe] shows no significant gradient as a function of $R_{gc}$ ($-0.003 \pm 0.012$ dex/kpc), as in Myers et al. (2022) ($0.004 \pm 0.004$ dex/kpc). Taken together, these results suggest that [K/Fe] does not show a strong gradient with $R_{gc}$. [K/H] shows a negative gradient with $R_{gc}$ ($-0.079 \pm 0.019$ dex/kpc), reflecting the disc's metallicity structure. Positive trends with [Fe/H] and age are also observed.

In summary, our findings for [Na/Fe], [Al/Fe], and [K/Fe] versus $R_{gc}$ are consistent with the majority of recent literature, which demonstrates either mild negative gradients or flat trends. Despite the overall abundance patterns being comparable, the [X/Fe] values reported by Myers et al. (2022) exhibit a noticeably higher degree of scatter when plotted against [Fe/H], stellar age, and $R_{gc}$, also for the odd-Z elements. In contrast, the data presented herein demonstrate more coherent trends and tighter groupings, likely a result of the more "classical" analysis-approach as compared to the industrial analysis of Myers et al. (2022). The chemical homogeneity of Al and K, in conjunction with the weak but potentially significant Na gradient, lends support to the hypothesis that these elements are well mixed in the Galactic thin disc, exhibiting only subtle variations that may be attributable to underlying disparities in stellar populations or the star formation history across the Galaxy.

### 4.5. Iron-peak elements (V, Cr, Mn, Co, and Ni)

The iron-peak elements are primarily formed in SNe Ia, which occur when a white dwarf in a binary system accretes material from its companion and eventually reaches a critical mass near the Chandrasekhar limit. This leads to a thermonuclear explosion that produces large amounts of iron (Fe) and other iron-peak elements (Whelan & Iben 1973; Iben & Tutukov 1984; Kobayashi et al. 2011, Kobayashi et al. 2020; Nomoto & Leung 2018). In this work, we analyzed V, Cr, Mn, Co, and Ni. Among these, Mn is particularly interesting, as its abundance trend is thought to depend on the nature and metallicity of the SNe Ia progenitors (de los Reyes et al. 2020). While SNe Ia are the dominant source of Mn, CCSNe may also contribute—particularly at lower metallicities. However, the behavior of Mn with respect to metallicity remains debated, with various studies presenting differing trends (e.g., Battistini & Bensby 2015; Zasowski et al. 2019; Lomaeva et al. 2019; Montelius et al. 2022; Vasini et al. 2024; Nandakumar et al. 2024b).

Vanadium: V abundances were derived using the H-band spectral line at 15924.84 Å. The stars UBC577_2, IC_4756_10, IC_4756_6, and UBC141_1 were excluded from the analysis due to the absence of detectable V lines, attributed to low S/N in the relevant spectral regions. A comparison of our V abundances with those of Myers et al. (2022) indicates broadly consistent results, with differences remaining within the reported uncertainties. The [V/Fe] ratios exhibit a slightly positive trend with $R_{gc}$, with a best-fit slope of $0.023 \pm 0.026$ dex/kpc, indicating no significant radial gradient. In contrast, [V/H] shows a slight negative trend with $R_{gc}$, with a slope of $-0.049 \pm 0.030$ dex/kpc, consistent with the general pattern of increasing V enrichment toward the inner Galaxy. Our results are consistent with the results of the OCCASO survey who reported a value of $0.001 \pm 0.003$ dex/kpc (Carbajo-Hijarrubia et al. 2024) for [V/Fe]. This is also consistent with the mild positive gradients and flat trend reported by Myers et al. (2022) and Magrini et al. (2023) respectively.

Chromium: Cr abundances were determined using two H-band spectral lines at 15680.06 Å and 15860.21 Å. A comparison of our Cr abundances with those of Myers et al. (2022) reveals consistent trends across the parameter space, with differences remaining within the uncertainties. The [Cr/Fe] ratios display a flat trend with [Fe/H], age, and $R_{gc}$, indicating a relatively uniform Cr-to-Fe production over time and across the disc. The best-fit slope for [Cr/Fe] versus $R_{gc}$ is $-0.004 \pm 0.010$ dex/kpc, while for [Cr/H] versus $R_{gc}$ it is $-0.076 \pm 0.022$ dex/kpc. This is broadly consistent, within the uncertainties, with several literature results. For example, Myers et al. (2022) give a slope for [Cr/Fe] to be $-0.002 \pm 0.005$ dex/kpc. Gaia-ESO reported a flat trend (Magrini et al. 2023). Carbajo-Hijarrubia et al. (2024) reported a value of $0.010 \pm 0.003$ dex/kpc. These mostly flat or weakly sloped trends further support the notion that Cr and Fe are produced in similar astrophysical sites, leading to well-coupled evolution across the disc.

Manganese: Mn abundances were derived using two H-band lines at 15217.74 Å and 15262.49 Å. NLTE synthesis was applied to both Mn lines used, but a LTE synthesis would render 0.01 dex higher [Mn/Fe] on average. The results were plotted and compared with those from Dal Ponte et al. (2025) and Myers et al. (2022), showing good agreement. The [Mn/Fe] ratios exhibit flat trend with $R_{gc}$, indicating a relatively consistent Mn-to-Fe production ratio across the Galactic disc. In contrast, [Mn/H] shows a clear downward trend with $R_{gc}$, reflecting progressive chemical enrichment over time. The best-fit slope for [Mn/Fe] vs. $R_{gc}$ is $-0.005 \pm 0.013$ dex/kpc, while for [Mn/H] vs. $R_{gc}$ it is $-0.077 \pm 0.020$ dex/kpc. Literature [Mn/Fe] values, such as Myers et al. (2022) ($-0.005 \pm 0.002$ dex/kpc), Magrini et al. (2023) ($0.003 \pm 0.010$ dex/kpc), and Carbajo-Hijarrubia et al. (2024) ($0.003 \pm 0.002$ dex/kpc) are consistent within uncertainties. The slight negative gradient we derive, suggests that Mn enrichment relative to Fe decreases marginally with increasing $R_{gc}$, possibly reflecting a metallicity-dependent nucleosynthetic origin for Mn or differences in Type Ia SNe contributions across the disc.

Cobalt: Co abundances were determined using the H-band spectral line at 16757.64 Å. The full stellar sample was analyzed, and the results were plotted with those from Dal Ponte et al. (2025) and Myers et al. (2022), showing strong agreement. The best-fit slope for [Co/Fe] vs. $R_{gc}$ is $0.024 \pm 0.011$ dex/kpc, while for [Co/H] versus $R_{gc}$ it is $-0.047 \pm 0.015$ dex/kpc, suggesting a mild decline in absolute Co abundances toward the outer disc. Our values are consistent within error bars with the small gradients found by Myers et al. (2022), Magrini et al. (2023), and Carbajo-Hijarrubia et al. (2024). The lack of a pronounced trend implies that Co and Fe are synthesized in similar proportions throughout the disc, resulting in a spatially uniform [Co/Fe] ratio.

Nickel: Ni abundances were determined using six H-band spectral lines as shown Table B.1. The full stellar sample was included in the analysis, and the results were plotted with those from Dal Ponte et al. (2025) and Myers et al. (2022), showing good overall agreement. The best-fit slope for [Ni/Fe] vs. $R_{gc}$ is $-0.019 \pm 0.012$ dex/kpc, while for [Ni/H] vs. $R_{gc}$ it is $-0.092 \pm 0.018$ dex/kpc, indicating a significant decrease in absolute Ni abundances toward the outer Galaxy. For [Ni/Fe], Myers et al. (2022) report a flatter value of $-0.004 \pm 0.002$ dex/kpc, while Magrini et al. (2023) also find a flat trend ($0.002 \pm 0.010$ dex/kpc). Carbajo-Hijarrubia et al. (2024) find a value of $-0.003 \pm 0.002$ dex/kpc. The negative gradient in our data, while still modest, could point to subtle variations in the nucleosynthetic processes that govern Ni and Fe production across the disc.

In summary, the trends for [V/Fe], [Cr/Fe], [Mn/Fe], [Co/Fe], and [Ni/Fe] as a function of $R_{gc}$ are mostly flat or show only weak gradients, in broad agreement with the bulk of recent literature. Small discrepancies and scatter between studies may be attributed to sample selection, measurement uncertainties, and the complex nucleosynthetic histories of the Fe-peak elements. Nevertheless, the overall chemical homogeneity of these ratios across the disc is clear, reinforcing the idea of efficient mixing in the interstellar medium and similar SNe enrichment histories for Fe-peak elements.

### 4.6. Neutron-capture elements (Cu, Zn, Y, Ce, Nd, and Yb)

As their name suggests, neutron-capture elements are formed through the capture of neutrons by atomic nuclei. This occurs via two primary processes distinguished by timescales: the slow (s-) process and the rapid (r-) process. The s-process occurs when neutron captures happen on timescales longer than the $\beta$-decay lifetimes of unstable nuclei. This process predominantly takes place in AGB stars and, to a lesser extent, in massive stars (Lugaro et al. 2023). In contrast, the r-process occurs in environments with a high density of free neutrons, allowing for rapid neutron captures before beta decay can occur. CCSNe and neutron star mergers (NSM) are the main sites proposed for the r-process (Cowan et al. 2021). In reality, most neutron-capture elements are produced by a combination of both s- and r-processes, though the dominant contributor varies by element.

Cu and Zn differ from the other investigated neutron-capture elements in that they are produced via the weak s-process in massive stars. Additionally, Cu differs from Zn in its nucleosynthetic origin: Cu is produced partly through the weak s-process in massive stars, but a significant fraction also originates in CCSNe (e.g., Pignatari et al. 2010). Zn, on the other hand, has only a minor weak s-process contribution; its production is dominated by explosive nucleosynthesis in CCSNe and hypernovae (e.g., Raiteri et al. 1991; Limongi & Chieffi 2003). This distinction underlines the role of Cu as a weak s-process element, whereas Zn is better described as a "transition element" bridging the iron-peak and neutron-capture regimes.

Y and Ce, on the other hand, are primarily produced via the main s-process in AGB-stars, while Nd is also largely of s-process origin, though with some r-process contribution (Bisterzo et al. 2014; Prantzos et al. 2020). Finally, Yb has a roughly 50-50 s- to r-process contribution, making it the element with the highest r-process fraction in this study (Bisterzo et al. 2014; Prantzos et al. 2020; Kobayashi et al. 2020).

**Copper:** Cu abundances were determined using two H-band lines at 16005.64 Å and 16006.44 Å. All sample stars were included in the analysis, and NLTE synthesis was applied to both Cu lines used, but a LTE synthesis would only render 0.01 dex lower [Cu/Fe] on average. The [Cu/Fe] ratios show a flat trend with $R_{gc}$, with a best-fit slope of $0.004 \pm 0.018$ dex/kpc. In contrast, [Cu/H] displays a negative gradient, with a best-fit slope of $-0.068 \pm 0.024$ dex/kpc, indicating increasing Cu enrichment toward the inner Galaxy. Gaia-ESO and OCCASO report a flat [Cu/Fe] trend and a slightly negative [Cu/H] trend which is consistent with our values (Magrini et al. 2023; Carbajo-Hijarrubia et al. 2024)

**Zinc:** In our work, Zn abundances were derived using a single H-band spectral line at 16505.18 Å. As only one line was used, the derived abundances are subject to larger uncertainties. Additionally, 47 stars were excluded from the Zn analysis due to poor line quality or unreliable measurements. The [Zn/Fe] and [Zn/H] ratios exhibit a declining trend with increasing [Fe/H], consistent with findings from previous studies such as Nandakumar et al. (2024b). In contrast, both [Zn/Fe] and [Zn/H] show positive gradients with $R_{gc}$, indicating higher Zn enrichment in the outer regions of the Galaxy. The best-fit slope for [Zn/Fe] versus $R_{gc}$ is $0.050 \pm 0.052$ dex/kpc, and for [Zn/H] versus $R_{gc}$ it is $-0.021 \pm 0.049$ dex/kpc.

**Yttrium:** Y abundances were derived using two IR lines at 21260.45 Å and 22543.84 Å. 44 stars were removed from the sample due to lines being too weak. The [Y/Fe] ratios exhibit a slight positive but statistically uncertain trend with $R_{gc}$, with a best-fit slope of $0.007 \pm 0.029$ dex/kpc. Meanwhile, [Y/H] shows a slightly negative distribution across $R_{gc}$, with a slope of $-0.006 \pm 0.038$ dex/kpc. Gaia-ESO reports an age-dependent trend in Y abundances, with [Y/Fe] increasing in younger clusters and [Y/H] exhibiting a negative slope with age. OCCASO reports a [Y/Fe] value of $0.011 \pm 0.004$ dex/kpc is in good agreement with our value.

**Cerium:** Ce abundances were determined using two H-band lines at 15977.12 Å and 16595.18 Å. The two stars UBC141_1 and NGC_2548_4 were excluded, as these Ce-lines were not detectable due to low S/N ratios. Our Ce abundances were compared with those from Myers et al. (2022), showing good agreement between the two datasets. The [Ce/Fe] ratios exhibit a mild positive trend with $R_{gc}$, with a best-fit slope of $0.021 \pm 0.028$ dex/kpc. In contrast, [Ce/H] shows a slight negative trend, with a slope of $-0.026 \pm 0.021$ dex/kpc, indicating a modest decline in absolute Ce abundances toward the outer Galaxy. This is also consistent with our previous studies, within the combined uncertainties (see; Bijavara Seshashayana et al. 2024a; Bijavara Seshashayana et al. 2024b). Our measured slope is consistent, within uncertainties, with the recent result by Myers et al. (2022), who find a slope of $0.022 \pm 0.006$ dex/kpc and by Magrini et al. (2009) (Gaia-ESO) who reported a flat trend for [Ce/Fe]. Carbajo-Hijarrubia et al. (2024) reports a value of $0.027 \pm 0.009$ for inner and $0.006 \pm 0.012$ for outer regions of the galaxy which is consistent with our value. These findings indicate that the [Ce/Fe] ratio does not vary strongly with $R_{gc}$, at least within the current measurement precision. The lack of a clear radial gradient could reflect the slow and widespread enrichment of the disc by the s-process, but it might also be due to the limited number of OCs with well-determined n-capture abundances.

**Neodynium:** Nd abundances were determined using four H-band lines as shown in the table B.1. The three stars NGC 2509_18, UPK85_2, and Berkeley_32_2 were excluded due to Nd II lines being too weak. The [Nd/Fe] ratios show a slight positive trend with $R_{gc}$, with a best-fit slope of $0.035 \pm 0.019$ dex/kpc. Conversely, [Nd/H] shows a negative trend with $R_{gc}$, with a slope of $-0.021 \pm 0.021$ dex/kpc, suggesting increasing Nd enrichment toward the inner regions of the Galaxy which is consistent with Gaia-ESO (Magrini et al. 2023). The [Nd/H] trends reported by Carbajo-Hijarrubia et al. (2024) are $0.049 \pm 0.009$ dex/kpc (inner) $0.016 \pm 0.016$ dex/kpc (outer) which is consistent within uncertainties when compared with our sample.

**Ytterbium:** Yb abundances were derived using the H-band line at 16498.40 Å. Several stars – UBC141_1, NGC

2509_18, Tombaugh_5_2, NGC_2548_4, NGC_2548_1, Berkeley_32_2, King11_7, and UBC577_2 – were excluded from the Yb analysis due to poor line quality or unreliable measurements. The [Yb/Fe] ratios show a positive trend with $R_{gc}$, with a best-fit slope of $0.065 \pm 0.031$ dex/kpc. In contrast, [Yb/H] decreases with $R_{gc}$, with a slope of $0.023 \pm 0.033$ dex/kpc, indicating a lesser Yb enrichment toward the inner Galaxy.

In summary, our results demonstrate that n-capture elements show only mild or negligible radial gradients across the Galactic disc, consistent with the most recent literature.

Investigating radial abundance gradients of n-capture elements in the Galactic disc remains challenging due to the relatively large uncertainties typically associated with their abundance determinations. In contrast to the well-studied $\alpha$- and Fe-peak elements, the distribution and evolution of n-capture elements hence remain less well understood. An additional source of uncertainty is the fact that these abundances are sensitive to a wide variety of astrophysical sites and timescales, including AGB stars and neutron star mergers (NSM), resulting in greater intrinsic scatter, which makes it difficult to establish robust Galactic trends. As more homogeneous and larger samples become available in the future, it may be possible to clarify whether subtle gradients exist for n-capture elements and what those trends reveal about the timescales and sources of their nucleosynthesis in the Milky Way.

## 5. Summary and conclusions

Trends of elemental abundances with Galactocentric radius reveal distinct behaviors across different nucleosynthetic groups, providing a window into the history and processes shaping the Galactic disc. Among the $\alpha$-elements, magnesium, silicon, calcium, and titanium exhibit essentially flat [X/Fe] trends as a function of $R_{gc}$, indicating a uniform enrichment relative to iron across the disc. Conversely, the [X/H] gradients exhibited by these star clusters are negative, thereby reflecting the well-established metallicity gradient that is characterised by a decrease in absolute abundances towards the outer regions of the Galaxy. Sulfur and fluorine demonstrate comparable behaviours; however, their [X/H] slopes are closer to zero or even slightly positive in some cases, suggesting the presence of additional complexity in their production channels or the possibility of sample effects.

In the case of the odd-Z elements, sodium and aluminium, there is evidence of flat or mildly negative [X/Fe] gradients and clear negative slopes in [X/H]. This finding indicates that while their relative production compared to iron remains stable, their absolute abundances are indicative of the overall metallicity decline with increasing $R_{gc}$. Potassium exhibits chemical homogeneity in both [X/Fe] and [X/H], with only marginal or statistically insignificant trends, consistent with its less certain nucleosynthetic origins.

Within the iron-peak group, elements such as, vanadium, chromium, manganese, cobalt, and nickel all display remarkably flat [X/Fe] profiles with respect to $R_{gc}$, thus reaffirming the hypothesis that their production is closely linked to that of iron itself – most likely through a combination of Type Ia and core-collapse supernovae.

Of particular interest is the observation that copper exhibits a flat [Cu/Fe] trend, yet a positive [Cu/H] gradient. This phenomenon may be indicative of local enrichment or the presence of distinct nucleosynthetic contributors. The neutron-capture elements (yttrium, cerium, neodymium, and ytterbium) have been observed to exhibit flat or mildly positive [X/Fe] gradients with $R_{gc}$, and typically flat or negative [X/H] slopes. These trends are likely to reflect the diverse nucleosynthetic sources for these species, such as the s-process in asymptotic giant branch stars and the r-process in neutron star mergers, where the former act over longer timescales than those for lighter elements. The trend of [Yb/Fe] has been explored for the first time, providing novel constraints.

The trans-iron element, zinc, is of particular interest, as it demonstrates positive slopes in both [Zn/Fe] and [Zn/H]. This suggests the potential for a distinct or more efficient production channel for Zn at larger $R_{gc}$, which may be linked to hypernovae or metallicity-dependent yields from massive stars.

A detailed comparison with values from the literature shows that, while our results are broadly consistent with recent optical and infrared studies within uncertainties, our analysis offers improved precision and homogeneity. This is especially evident for elements like Fluorine and Potassium that are challenging to measure in the optical domain, where our high-resolution infrared approach provides more reliable abundance determinations. The enhanced internal consistency and reduced scatter across the dataset demonstrate the robustness of the methodology and underscore the advantages of a uniform, infrared-based analysis for studying open cluster chemistry.

Finally, this work highlights the critical role of open clusters as tracers of Galactic chemical evolution. As young, coeval stellar populations with accurately determined ages, distances and kinematics, open clusters are ideal for investigating element abundance patterns across a wide range of Galactic environments. Large open cluster-based samples, which covers a broad spectrum of $R_{gc}$, metallicity, and age, enables the identification of subtle abundance gradients and the disentangling of nucleosynthetic signatures. It also allows for a nuanced assessment of the impact of different enrichment processes, including Type II and Ia supernovae, AGB stars, and neutron-capture events, on the evolution of the Milky Way. Furthermore, this study fills an important gap in the chemical cartography of the Galaxy by extending the analysis to more elements being analyzed consistently within one project, including many key odd-Z and neutron-capture elements. Our results therefore provide valuable and unique insights, furthering our understanding of the complex interplay of processes that drive Galactic chemical evolution.

*Acknowledgements.* Based on observations with the Italian Telescopio Nazionale Galileo (TNG) operated on the island of La Palma by the Fundación Galileo Galilei of the INAF (Istituto Nazionale di Astrofisica) at the Spanish Observatorio del Roque de los Muchachos of the Instituto de Astrofísica de Canarias. We thank the anonymous referee for their helpful comments and suggestions that improved this paper. S.B.S. acknowledges funding from the Crafoord Foundation. H.J. acknowledges support from the Swedish Research Council, VR (grant 2024-04989). AB acknowledges support from INAF MiniGrant 2022. V.D. acknowledges financial support from the Fulbright Visiting Research Scholar 2024-2025.

Table 1: Comparison of Galactic abundance gradients (slopes in dex/kpc) for [X/H] vs. $R_{gc}$ from this work and selected literature sources.

| Element | This work | Myers et al. (2022) | OCCASO | Gaia-ESO |
|---------|-----------|---------------------|--------|----------|
| Fe | $-0.072 \pm 0.017$ | $-0.073 \pm 0.002$ | $-0.059 \pm 0.017$ | $-0.054 \pm 0.004$ |
| F | $-0.118 \pm 0.060$ | — | — | — |
| Mg | $-0.061 \pm 0.016$ | $-0.027 \pm 0.006$ | $-0.042 \pm 0.020$ | $-0.058 \pm 0.006$ |
| Si | $-0.044 \pm 0.016$ | $-0.031 \pm 0.006$ | $-0.046 \pm 0.018$ | $-0.039 \pm 0.004$ |
| S | $-0.007 \pm 0.023$ | $-0.028 \pm 0.006$ | — | — |
| Ca | $-0.069 \pm 0.017$ | $-0.031 \pm 0.006$ | $-0.046 \pm 0.018$ | $-0.034 \pm 0.003$ |
| Na | $-0.113 \pm 0.028$ | $-0.039 \pm 0.010$ | $-0.086 \pm 0.019$ | $-0.061 \pm 0.005$ |
| Al | $-0.069 \pm 0.018$ | $-0.028 \pm 0.007$ | $-0.072 \pm 0.018$ | $-0.046 \pm 0.007$ |
| K | $-0.079 \pm 0.019$ | $-0.027 \pm 0.007$ | — | — |
| Ti | $-0.069 \pm 0.018$ | $-0.034 \pm 0.007$ | $-0.043 \pm 0.018$ | $-0.045 \pm 0.005$ |
| V | $-0.049 \pm 0.030$ | $0.003 \pm 0.011$ | $-0.069 \pm 0.018$ | $-0.044 \pm 0.005$ |
| Cr | $-0.076 \pm 0.022$ | $-0.042 \pm 0.008$ | $-0.042 \pm 0.019$ | $-0.053 \pm 0.005$ |
| Mn | $-0.077 \pm 0.020$ | $-0.040 \pm 0.007$ | $-0.066 \pm 0.020$ | $-0.061 \pm 0.007$ |
| Co | $-0.047 \pm 0.015$ | $-0.038 \pm 0.011$ | $-0.066 \pm 0.018$ | $-0.059 \pm 0.006$ |
| Ni | $-0.092 \pm 0.018$ | $-0.032 \pm 0.007$ | $-0.059 \pm 0.017$ | $-0.053 \pm 0.005$ |
| Cu | $-0.068 \pm 0.024$ | — | $-0.064 \pm 0.020$ | $-0.056 \pm 0.010$ |
| Zn | $-0.021 \pm 0.049$ | — | $-0.033 \pm 0.021$ | $-0.046 \pm 0.004$ |
| Y | $-0.006 \pm 0.038$ | — | $-0.064 \pm 0.020$ | $-0.045 \pm 0.004$ |
| Ce | $-0.026 \pm 0.021$ | $-0.006 \pm 0.009$ | $-0.058 \pm 0.021$ | $-0.037 \pm 0.004$ |
| Nd | $-0.021 \pm 0.021$ | — | $-0.027 \pm 0.023$ | $-0.015 \pm 0.003$ |
| Yb | $0.023 \pm 0.033$ | — | — | — |

**Notes.** The literature values are taken from Myers et al. (2022), Carbajo-Hijarrubia et al. (2024), and Magrini et al. (2023). APOGEE and GALAH surveys cover a larger $R_{gc}$ interval, while OCCASO spans a range comparable to that used in this work.

## Data availability

All machine readable data underlying this work are available in electronic form at the CDS via anonymous ftp to `cdsarc.u-strasbg.fr` (130.79.128.5) or via `http://cdsweb.u-strasbg.fr/cgi-bin/qcat?J/A+A/`. The CDS package includes: (i) the full list of observed stars and basic properties for each open cluster (Appendix **??**); (ii) the complete atomic and molecular line list used in the analysis (Appendix **??**); (iii) derived stellar parameters with formal PySME uncertainties and a literature-comparison table discussed (Appendix **??**); and (iv) per-star abundance ratios [X/Fe] with uncertainties, cluster means and dispersions, and per-element star counts documenting sample completeness (Appendix **??**). All files include column descriptions and are provided in machine-readable ASCII format.

# Appendix A: Overview of observed clusters and stars

Table A.1: The following table provides general information for the stars analysed in each stellar cluster, including the Gaia DR3 source ID, equatorial coordinates, and *G*-band magnitude. The ages, A$_V$, and R$_{gc}$ are from Cantat-Gaudin et al. (2020) for all clusters except for NGC 6791 which is from Brogaard et al. (2021). The clusters are arranged in ascending order of age. Only a segment of the table is presented here and the complete table can be accessed at the CDS.

| Stellar cluster | Age (Gyr) | R$_{gc}$ (kpc) | A$_V$ (mag) | Star ID | *Gaia* DR3 ID | RA (deg) | Dec (deg) | G (mag) |
|---|---|---|---|---|---|---|---|---|
| Gulliver 18 | 0.04 | 7.82 | 1.59 | Gulliver__18 | 1836389309820904064 | 302.9329000 | 26.5852740 | 8.96 |
| Collinder 463 | 0.11 | 8.87 | 0.79 | Collinder__463_1 | 534346643760560896 | 26.7492357 | 71.7603484 | 7.73 |
| | | | | Collinder__463_2 | 534346643760560896 | 23.4560360 | 71.8526614 | 8.02 |
| | | | | Collinder__463_3 | 534363067715447680 | 26.2881410 | 71.8903579 | 7.95 |
| Alessi Teutch 11 | 0.14 | 8.33 | 0.37 | Alessi__Teutch__11 | 2184332753719499904 | 304.0934800 | 52.1051250 | 7.11 |
| UPK 219 | 0.15 | 8.74 | 1.2 | UPK__219 | 2209440823287736064 | 352.6238249 | 65.1431365 | 8.73 |
| Gulliver 24 | 0.18 | 9.13 | 1.05 | Gulliver__24 | 430035249779499264 | 1.1186123 | 62.7012180 | 10.50 |
| Tombaugh 5 | 0.19 | 9.77 | 2.07 | Tombaugh__5_2 | 473275782228263296 | 57.1373856 | 59.2545867 | 11.17 |
| | | | | Tombaugh__5_4 | 473266779976916480 | 56.8791316 | 59.0474543 | 11.37 |
| ... | ... | ... | ... | ... | ... | ... | ... | ... |
| ... | ... | ... | ... | ... | ... | ... | ... | ... |
| ... | ... | ... | ... | ... | ... | ... | ... | ... |

## Appendix B: Atomic and Molecular data

Table B.1: Spectral line data of molecular (OH, CO, CN, HF) and atomic (Fe I, Mg I, Si I, Ti I/II, S I, Ca I, Na I, Al I, K I, V I, Cr I, Mn I, Co I, Ni I, Cu I, Zn I, Y I, Ce II, Nd II, Yb I) lines used in determining stellar parameters and elemental abundances. $\log(gf)$-values are from Brooke et al. (2016) for OH, Li et al. (2015) for CO, Sneden et al. (2014) for CN. For all other lines, we used updated VALD data (Piskunov et al. 1995; Kupka et al. 2000; Ryabchikova et al. 2015) with astrophysical calibrations based on solar and Arcturus spectra (Montelius et al. 2022; Nandakumar et al. 2023a,b, 2024a). Only a segment of the table is presented here and the complete table can be accessed at the CDS.

| Species | $\lambda_{\mathrm{air}}$ (Å) | $\log(gf)$ (dex) | $E_{\mathrm{low}}$ (eV) |
|---|---|---|---|
| OH | 15222.114 | -7.336 | 2.686 |
| | 15409.168 | -5.435 | 0.255 |
| | ... | ... | ... |
| CO | 15585.352 | -5.872 | 5.351 |
| | 15689.155 | -5.839 | 4.579 |
| | ... | ... | ... |
| CN | 15195.281 | -4.031 | 0.671 |
| | 15206.737 | -5.768 | 2.374 |
| | ... | ... | ... |
| HF | 23358.33 | -3.962 | 0.227 |
| Fe I | 15160.101 | -2.170 | 2.627 |
| | 15160.101 | -2.170 | 6.342 |
| | ... | ... | ... |
| Mg I | 15231.593 | -1.579 | 6.719 |
| | 15693.311 | -1.200 | 6.719 |
| | ... | ... | ... |
| Si I | 15376.939 | -2.397 | 6.721 |
| | 15884.454 | -0.819 | 5.954 |
| | ... | ... | ... |
| Ti I | 15230.629 | -5.750 | 4.396 |
| | 15316.472 | -3.405 | 5.191 |
| | ... | ... | ... |
| Ti II | 15873.843 | -1.997 | 3.123 |
| S I | 15478.482 | -0.047 | 8.046 |
| Ca I | 16150.750 | -0.244 | 4.532 |
| | 16155.244 | -0.674 | 4.532 |
| | ... | ... | ... |
| Na I | 16373.864 | -1.765 | 3.753 |
| | 16388.862 | -2.590 | 3.753 |
| | ... | ... | ... |
| Al I | 16718.925 | -0.369 | 4.085 |
| | 16750.550 | -0.668 | 4.087 |
| | ... | ... | ... |
| K I | 15168.376 | 0.378 | 2.670 |
| V I | 15924.821 | -1.096 | 2.138 |
| Cr I | 15680.060 | 0.180 | 4.697 |
| | 15860.210 | 0.061 | 4.697 |
| Mn I | 15217.715 | -0.057 | 4.889 |
| | 15262.459 | -0.134 | 4.889 |
| Co I | 16757.613 | -2.181 | 3.409 |
| Ni I | 16310.501 | -0.024 | 5.283 |
| | 16363.103 | 0.412 | 5.283 |
| | ... | ... | ... |
| Cu I | 16005.637 | -0.466 | 5.348 |
| | 16005.647 | -1.516 | 5.348 |
| | ... | ... | ... |
| Zn I | 16505.180 | 0.751 | 7.783 |
| Y I | 21260.450 | -0.608 | 1.398 |
| | 22543.840 | -0.767 | 1.429 |
| Ce II | 15977.120 | -2.100 | 0.232 |
| | 16595.180 | -2.144 | 0.122 |
| Nd II | 15368.153 | -1.550 | 1.264 |
| | 16053.628 | -2.200 | 0.744 |
| | ... | ... | ... |
| Yb I | 16498.400 | -0.640 | 3.017 |

# Appendix C: Stellar parameters

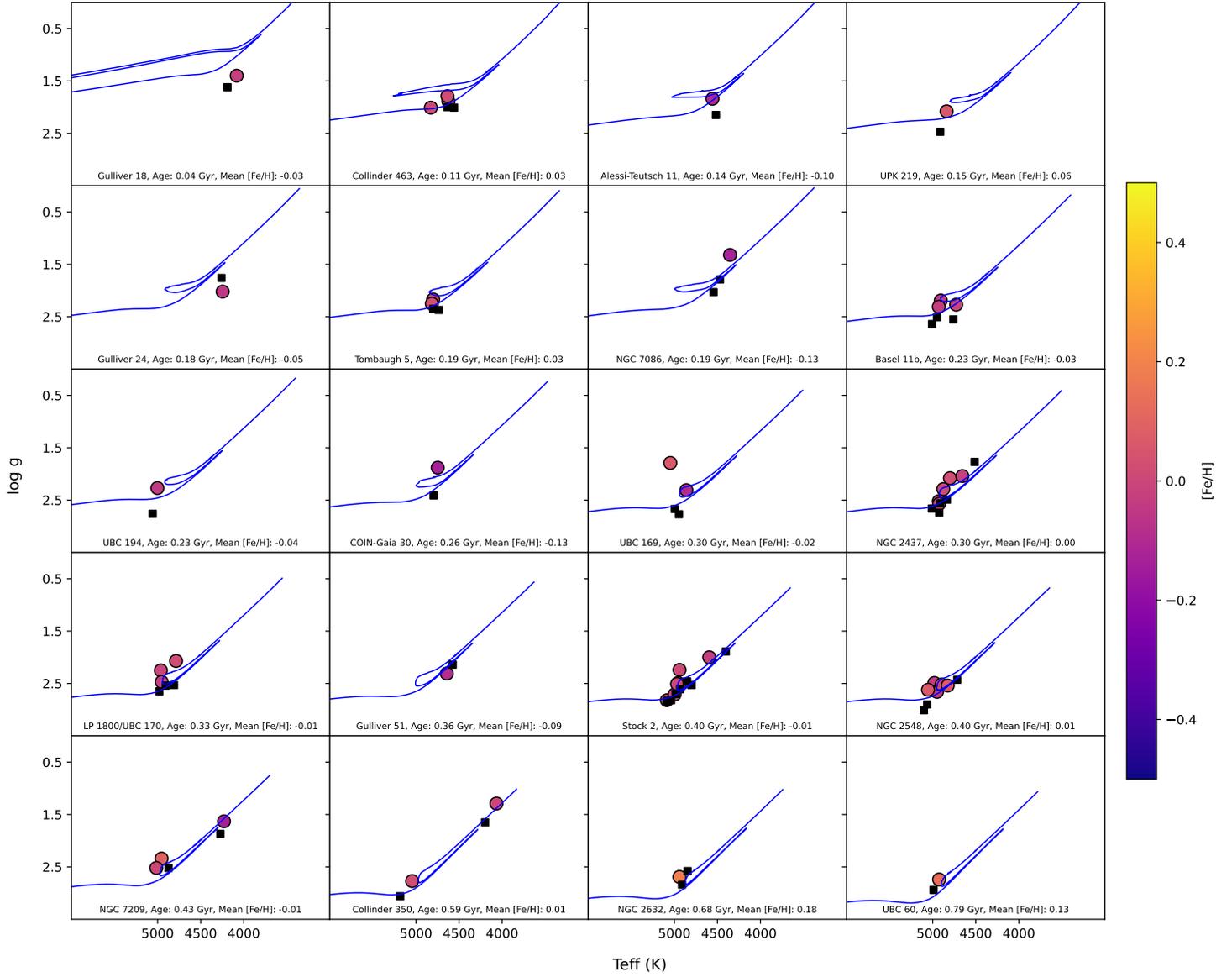

Fig. C.1: Kiel diagrams showing $T_{eff}$, $\log g$ and [Fe/H] of stars belonging to different clusters together with the corresponding isochrone from the MIST database. Each panel corresponds to a different OC, with its name, age and our mean [Fe/H] stated in the plot. Optical data from Dal Ponte et al. (2025) are overplotted as black dots. The dataset incorporates all stars from both their study and our analysis, without restricting the selection to those in common. Notes. Isochrones and the data points do not always match perfectly, reflecting uncertainties in both stellar models and cluster parameters.

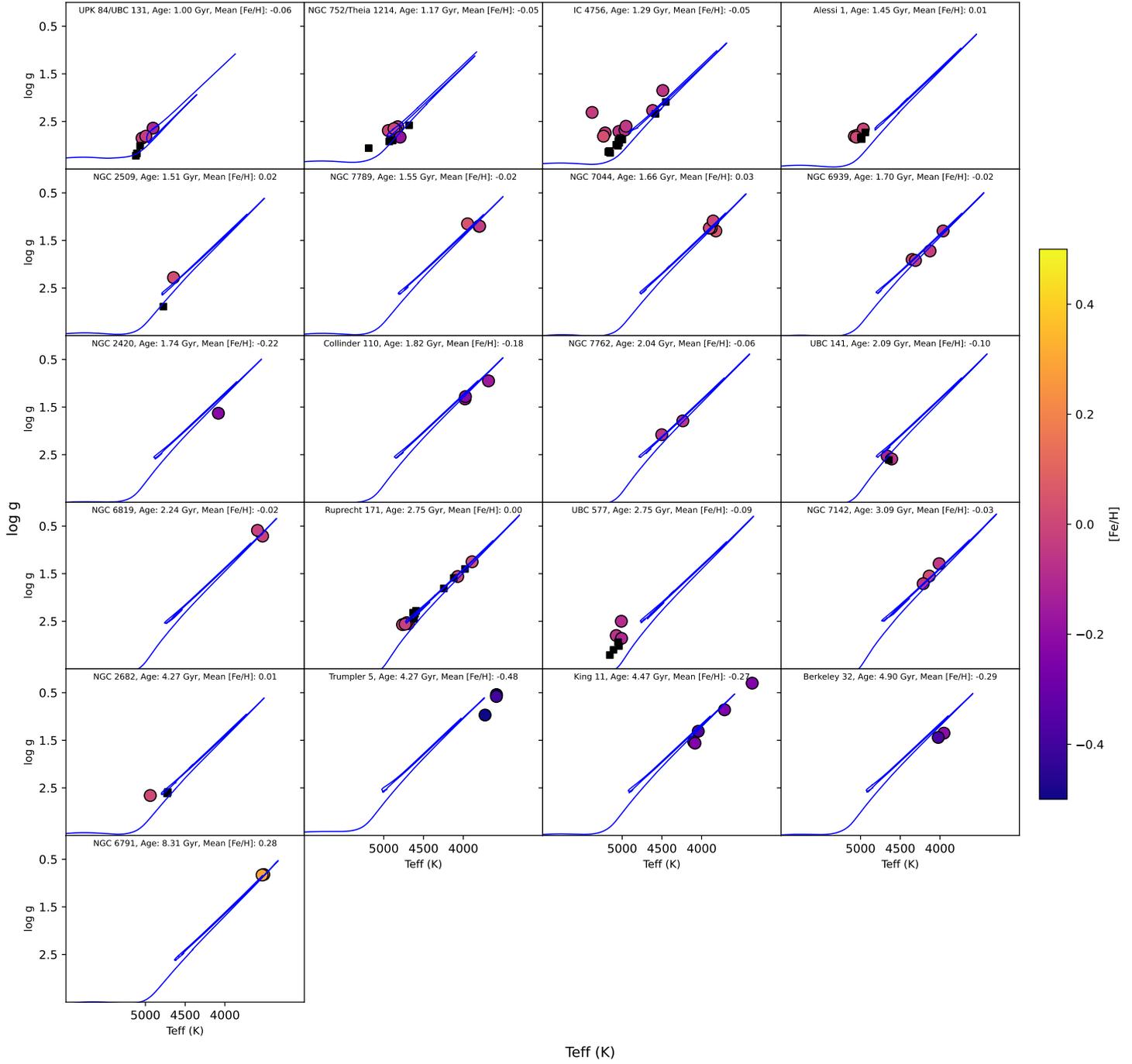

Fig. C.2: Same as Figure C.1, but for the remaining clusters.

Table C.1: Stellar parameters $T_{eff}$, $\log g$, [Fe/H], $v_{mic}$, $v_{mac}$, [C/Fe], [N/Fe], and [O/Fe] for the sample stars in this work along with their formal PySME fit uncertainties. Only a segment of the table is presented here and the complete table can be accessed at the CDS.

| Cluster | Star | $T_{eff}$ (K) | $\log g$ (dex) | [Fe/H] (dex) | $v_{mic}$ (km s$^{-1}$) | $v_{mac}$ (km s$^{-1}$) | [C/Fe] (dex) | [N/Fe] (dex) | [O/Fe] (dex) |
|---|---|---|---|---|---|---|---|---|---|
| Gulliver 18 | Gulliver_18 | 4082 ± 16 | 1.40 ± 0.007 | −0.03 ± 0.01 | 2.15 ± 0.05 | 6.60 ± 0.13 | −0.04 ± 0.03 | 0.44 ± 0.03 | 0.07 ± 0.02 |
| Collinder 463 | Collinder_463_1 | 4623 ± 21 | 1.89 ± 0.004 | 0.06 ± 0.01 | 2.00 ± 0.07 | 7.75 ± 0.14 | −0.21 ± 0.03 | 0.59 ± 0.04 | 0.12 ± 0.03 |
| | Collinder_463_2 | 4827 ± 4 | 2.00 ± 0.001 | 0.01 ± 0.02 | 2.25 ± 0.07 | 12.53 ± 0.18 | −0.23 ± 0.04 | 0.50 ± 0.05 | 0.08 ± 0.02 |
| | Collinder_463_3 | 4635 ± 26 | 1.79 ± 0.005 | 0.02 ± 0.02 | 2.06 ± 0.08 | 8.58 ± 0.17 | −0.20 ± 0.03 | 0.56 ± 0.04 | 0.10 ± 0.03 |
| Alessi Teutsch 11 | Alessi_Teutsch_11 | 4557 ± 22 | 1.84 ± 0.004 | −0.10 ± 0.01 | 1.86 ± 0.07 | 7.30 ± 0.16 | −0.24 ± 0.03 | 0.54 ± 0.04 | 0.12 ± 0.03 |
| UPK 219 | UPK_219 | 4839 ± 2 | 2.08 ± 0.001 | 0.06 ± 0.02 | 1.89 ± 0.07 | 9.34 ± 0.16 | −0.17 ± 0.03 | 0.39 ± 0.05 | 0.06 ± 0.03 |
| Gulliver 24 | Gulliver_24 | 4245 ± 19 | 2.02 ± 0.004 | −0.05 ± 0.01 | 1.60 ± 0.06 | 5.98 ± 0.16 | −0.17 ± 0.03 | 0.44 ± 0.04 | 0.03 ± 0.03 |
| Tombaugh 5 | Tombaugh_5_2 | 4815 ± 3 | 2.25 ± 0.001 | 0.03 ± 0.02 | 1.97 ± 0.09 | 7.72 ± 0.20 | −0.13 ± 0.05 | 0.51 ± 0.07 | 0.07 ± 0.03 |
| | Tombaugh_5_4 | 4800 ± 27 | 2.17 ± 0.005 | 0.03 ± 0.02 | 2.00 ± 0.07 | 11.32 ± 0.16 | −0.28 ± 0.04 | 0.66 ± 0.05 | 0.10 ± 0.02 |
| ... | ... | ... | ... | ... | ... | ... | ... | ... | ... |
| ... | ... | ... | ... | ... | ... | ... | ... | ... | ... |
| ... | ... | ... | ... | ... | ... | ... | ... | ... | ... |

Table C.2: Comparison of the stellar parameters T$_{eff}$, log $g$ and [Fe/H] for the sample stars in this work and literature values for the same stars. Differences are (this work − literature). Uncertainties are standard deviations for each cluster.

| Cluster | No. of stars in common | T$_{eff}$ (K) | log $g$ (dex) | [Fe/H] | Reference |
|---|---|---|---|---|---|
| Gulliver 18 | 1 | −108 | −0.22 | −0.02 | Dal Ponte et al. (2025) |
| | 1 | −517 | −0.10 | −0.28 | Zhang et al. (2021) |
| Collinder 463 | 2 | 33 ± 46 | −0.16 ± 0.07 | 0.10 ± 0.01 | Dal Ponte et al. (2025) |
| | 1 | −5 ± 144 | −0.26 ± 0.21 | −0.22 ± 0.04 | Zhang et al. (2021) |
| Alessi Teutsch 11 | 1 | 40 | −0.30 | −0.06 | Dal Ponte et al. (2025) |
| | 1 | −233 | −1.56 | −0.20 | Zhang et al. (2021) |
| UPK 219 | 1 | −74 | −0.39 | −0.01 | Dal Ponte et al. (2025) |
| | 1 | −364 | −0.93 | −0.19 | Zhang et al. (2021) |
| Gulliver 24 | 1 | −14 | 0.25 | 0.09 | Dal Ponte et al. (2025) |
| | 1 | −322 | 0.39 | −0.30 | Zhang et al. (2021) |
| Tombaugh 5 | 2 | 40 ± 35 | −0.15 ± 0.07 | 0.04 ± 0.09 | Dal Ponte et al. (2025) |
| | 1 | −206 | −0.15 | −0.03 | Zhang et al. (2021) |
| NGC 7086 | 1 | −117 | −0.47 | −0.03 | Dal Ponte et al. (2025) |
| Basel 11 b | 3 | −52 ± 24 | −0.31 ± 0.03 | −0.06 ± 0.03 | Dal Ponte et al. (2025) |
| | 3 | −1102 ± 218 | 0.55 ± 0.15 | −0.04 ± 0.01 | Zhang et al. (2021) |
| UBC 194 | 1 | −55 | −0.49 | −0.10 | Dal Ponte et al. (2025) |
| COIN Gaia 30 | 1 | −48 | −0.52 | −0.09 | Dal Ponte et al. (2025) |
| | 1 | −483 | 0.18 | −0.38 | Zhang et al. (2021) |
| UBC 169 | 3 | −18 ± 98 | −0.67 ± 0.29 | −0.08 ± 0.08 | Dal Ponte et al. (2025) |
| NGC 2437 | 4 | 12 ± 99 | 0.04 ± 0.20 | 0.01 ± 0.09 | Dal Ponte et al. (2025) |
| | 3 | −104 ± 105 | 0.09 ± 0.06 | −0.04 ± 0.01 | Zhang et al. (2021) |
| UBC 170 | 3 | 3 ± 176 | −0.31 ± 0.26 | −0.10 ± 0.04 | Dal Ponte et al. (2025) |
| Gulliver 51 | 1 | 68 | 0.17 | 0.08 | Dal Ponte et al. (2025) |
| | 1 | −89 | −0.14 | −0.35 | Casali et al. (2020) [FAMA] |
| | 1 | 1 | −0.35 | −0.09 | Casali et al. (2020) [ROTFIT] |
| Stock 2 | 5 | −43 ± 113 | −0.21 ± 0.23 | 0.02 ± 0.05 | Dal Ponte et al. (2025) |
| | 7 | −47 ± 199 | −0.43 ± 0.28 | −0.01 ± 0.03 | Alonso-Santiago et al. (2021) |
| NGC 2548 | 3 | −16 ± 131 | −0.17 ± 0.25 | −0.01 ± 0.09 | Dal Ponte et al. (2025) |
| NGC 7209 | 2 | 19 ± 87 | −0.21 ± 0.04 | 0.00 ± 0.07 | Dal Ponte et al. (2025) |
| | 2 | 121 ± 977 | −0.02 ± 1.21 | −0.04 ± 0.16 | Zhang et al. (2021) |
| Collinder 350 | 2 | −135 ± 5 | −0.32 ± 0.05 | 0.04 ± 0.15 | Dal Ponte et al. (2025) |
| | 2 | −195 ± 85 | −0.22 ± 0.30 | −0.09 ± 0.01 | Zhang et al. (2021) |
| | 2 | −80 ± 64 | −0.07 ± 0.02 | −0.12 ± 0.16 | Casali et al. (2020) [FAMA] |
| | 2 | −145 ± 170 | −0.10 ± 0.17 | −0.01 ± 0.02 | Casali et al. (2020) [ROTFIT] |
| NGC 2632 | 1 | 30 | −0.15 | −0.02 | Dal Ponte et al. (2025) |
| UBC 60 | 1 | −65 | −0.19 | −0.10 | Dal Ponte et al. (2025) |
| UPK 84 | 3 | −117 ± 50 | −0.37 ± 0.04 | −0.11 ± 0.02 | Dal Ponte et al. (2025) |

| Cluster | No. of stars in common | T$_{eff}$ (K) | log $g$ (dex) | [Fe/H] | Reference |
|---|---|---|---|---|---|
| NGC 752 | 4 | $-29 \pm 101$ | $-0.13 \pm 0.26$ | $-0.06 \pm 0.09$ | Dal Ponte et al. (2025) |
| | 1 | $-36 \pm 0.00$ | $-0.02 \pm 0.00$ | $-0.04 \pm 0.00$ | Casamiquela et al. 2017, 2019 |
| | 1 | $-257 \pm 0.00$ | $-0.40 \pm 0.00$ | $-0.14 \pm 0.00$ | Carlos et al. (2018) |
| | 2 | $-112 \pm 141$ | $-0.22 \pm 0.16$ | $0.04 \pm 0.07$ | Böcek Topcu et al. (2015) |
| | 4 | $-71 \pm 116$ | $-0.03 \pm 0.19$ | $0.09 \pm 0.05$ | Reddy et al. (2012) |
| | 1 | $-58$ | $-0.32$ | $0.05$ | Carrera & Pancino (2011) |
| IC 4756 | 6 | $43 \pm 139$ | $-0.31 \pm 0.18$ | $-0.04 \pm 0.05$ | Dal Ponte et al. (2025) |
| | 4 | $74 \pm 56$ | $-0.08 \pm 0.10$ | $-0.07 \pm 0.03$ | Casamiquela et al. 2017, 2019 |
| | 3 | $-162 \pm 311$ | $-0.11 \pm 0.83$ | $-0.25 \pm 0.01$ | Jacobson et al. (2007) |
| | 3 | $-380 \pm 299$ | $-0.70 \pm 0.45$ | $-0.11 \pm 0.02$ | Santos et al. (2009) |
| | 2 | $-207 \pm 26$ | $-0.47 \pm 0.06$ | $-0.19 \pm 0.00$ | Pace et al. (2010) |
| Alessi 1 | 4 | $58 \pm 28$ | $-0.05 \pm 0.02$ | $0.02 \pm 0.01$ | Dal Ponte et al. (2025) |
| | 4 | $358 \pm 169$ | $-0.09 \pm 0.88$ | $-0.23 \pm 0.02$ | Zhang et al. (2021) |
| NGC 2509 | 1 | $-128$ | $-0.61$ | $-0.19$ | Dal Ponte et al. (2025) |
| | 1 | $-59$ | $0.01$ | $0.02$ | Zhang et al. (2021) |
| NGC 7789 | 1 | $98.5$ | $0.60$ | $0.83$ | García Pérez et al. (2021) |
| NGC 7044 | 4 | $-86 \pm 55$ | $-0.04 \pm 0.04$ | $0.21 \pm 0.07$ | Dal Ponte et al. (2025) |
| NGC 6939 | 1 | $9$ | $0.22$ | $-0.06$ | Soubiran et al. (2022) |
| NGC 2420 | 1 | $-241$ | $-0.17$ | $0.37$ | Soubiran et al. (2022) |
| Collinder 110 | - | - | - | - | - |
| NGC 7762 | 1 | $288$ | $0.25$ | $-1.42$ | Carraro et al. (2006) |
| | 4 | $-118 \pm 36$ | $0.09 \pm 0.13$ | $-0.34 \pm 0.03$ | Casali et al. (2020) [FAMA] |
| | 4 | $-58 \pm 30$ | $-0.29 \pm 0.09$ | $-0.10 \pm 0.01$ | Casali et al. (2020) [ROTFIT] |
| UBC 141 | 1 | $21$ | $-0.08$ | $-0.10$ | Dal Ponte et al. (2025) |
| NGC 6819 | 2 | $-250 \pm 61$ | $-0.44 \pm 0.09$ | $0.50 \pm 0.32$ | Lee-Brown et al. (2015) |
| Ruprecht 171 | 5 | $51 \pm 99$ | $0.14 \pm 0.15$ | $0.08 \pm 0.07$ | Dal Ponte et al. (2025) |
| | 6 | $-97 \pm 45$ | $-0.16 \pm 0.18$ | $-0.14 \pm 0.12$ | Casali et al. (2020) [FAMA] |
| | 6 | $52 \pm 59$ | $-0.10 \pm 0.11$ | $-0.10 \pm 0.02$ | Casali et al. (2020) [ROTFIT] |
| UBC 577 | 3 | $-51 \pm 26$ | $-0.34 \pm 0.16$ | $-0.05 \pm 0.03$ | Dal Ponte et al. (2025) |
| NGC 7142 | 1 | $-189$ | $-0.19$ | $-0.19$ | Soubiran et al. (2022) |
| NGC 2682 | 1 | $338$ | $0.16$ | $0.01$ | Zhang et al. (2021) |
| Trumpler 5 | 1 | $-150$ | $0.27$ | $-0.01$ | Özdemir et al. (2025) |
| King 11 | - | - | - | - | - |
| Berkeley 32 | 2 | $-273 \pm 100$ | $-0.23 \pm 0.29$ | $0.04 \pm 0.27$ | Hourihane et al. (2023) |
| NGC 6791 | 1 | $-18$ | $0.06$ | $0.02$ | Abdurro'uf et al. (2022) |

## Appendix D: Abundances

Table D.1: The number of stars in each cluster for which a given element was possible to determine. Individual stars were excluded if the spectral line was not detected or of poor quality (low S/N). The final row provides the total number of stars per element across all clusters.

| Stellar cluster | Total no of stars | C | N | F | Mg | Si | S | Ca | Na | Al | K | Ti | V | Cr | Mn | Co | Ni | Cu | Zn | Y | Ce | Nd | Yb |
|---|---|---|---|---|---|---|---|---|---|---|---|---|---|---|---|---|---|---|---|---|---|---|---|
| Gulliver 18 | 1 | 1 | 1 | 1 | 1 | 1 | 1 | 1 | 1 | 1 | 1 | 1 | 1 | 1 | 1 | 1 | 1 | 1 | 0 | 1 | 1 | 1 | 1 |
| Collinder 463 | 3 | 3 | 3 | 0 | 3 | 3 | 3 | 3 | 3 | 3 | 3 | 3 | 3 | 3 | 3 | 3 | 3 | 3 | 0 | 1 | 3 | 3 | 3 |
| Alessi Teutch 11 | 1 | 1 | 1 | 0 | 1 | 1 | 1 | 1 | 1 | 1 | 1 | 1 | 1 | 1 | 1 | 1 | 1 | 1 | 1 | 1 | 1 | 1 | 1 |
| UPK 219 | 1 | 1 | 1 | 0 | 1 | 1 | 1 | 1 | 1 | 1 | 1 | 1 | 1 | 1 | 1 | 1 | 1 | 1 | 0 | 1 | 1 | 1 | 1 |
| Gulliver 24 | 1 | 1 | 1 | 0 | 1 | 1 | 1 | 1 | 1 | 1 | 1 | 1 | 1 | 1 | 1 | 1 | 1 | 1 | 1 | 1 | 1 | 1 | 1 |
| Tombaugh 5 | 2 | 2 | 2 | 0 | 2 | 2 | 2 | 2 | 2 | 2 | 2 | 2 | 2 | 2 | 2 | 2 | 2 | 2 | 1 | 1 | 2 | 2 | 1 |
| NGC 7086 | 1 | 1 | 1 | 0 | 1 | 1 | 1 | 1 | 1 | 1 | 1 | 1 | 1 | 1 | 1 | 1 | 1 | 1 | 1 | 1 | 1 | 1 | 1 |
| Basel 11b | 3 | 3 | 3 | 0 | 3 | 3 | 3 | 3 | 3 | 3 | 3 | 3 | 3 | 3 | 3 | 3 | 3 | 3 | 0 | 1 | 3 | 3 | 3 |
| UBC 194 | 1 | 1 | 1 | 0 | 1 | 1 | 1 | 1 | 1 | 1 | 1 | 1 | 1 | 1 | 1 | 1 | 1 | 1 | 1 | 0 | 1 | 1 | 1 |
| COIN-Gaia 30 | 1 | 1 | 1 | 0 | 1 | 1 | 1 | 1 | 1 | 1 | 1 | 1 | 1 | 1 | 1 | 1 | 1 | 1 | 1 | 1 | 1 | 1 | 1 |
| UBC 169 | 2 | 2 | 2 | 0 | 2 | 2 | 2 | 2 | 2 | 2 | 2 | 2 | 2 | 2 | 2 | 2 | 2 | 2 | 2 | 2 | 2 | 2 | 2 |
| NGC 2437 | 5 | 5 | 5 | 0 | 5 | 5 | 5 | 5 | 5 | 5 | 5 | 5 | 5 | 5 | 5 | 5 | 5 | 5 | 5 | 3 | 4 | 5 | 5 |
| LP 1800 | 5 | 3 | 3 | 0 | 3 | 3 | 3 | 3 | 3 | 3 | 3 | 3 | 3 | 3 | 3 | 3 | 3 | 3 | 2 | 1 | 3 | 3 | 3 |
| Gulliver 51 | 1 | 1 | 1 | 0 | 1 | 1 | 1 | 1 | 1 | 1 | 1 | 0 | 1 | 1 | 1 | 1 | 1 | 1 | 1 | 1 | 1 | 1 | 1 |
| Stock 2 | 7 | 7 | 7 | 0 | 7 | 7 | 7 | 7 | 7 | 7 | 7 | 7 | 7 | 7 | 7 | 7 | 7 | 7 | 0 | 3 | 7 | 7 | 7 |
| NGC 2548 | 5 | 5 | 5 | 0 | 5 | 5 | 5 | 5 | 5 | 5 | 5 | 5 | 5 | 5 | 5 | 5 | 5 | 5 | 0 | 2 | 4 | 5 | 3 |
| NGC 7209 | 3 | 3 | 3 | 0 | 3 | 3 | 3 | 3 | 3 | 3 | 3 | 3 | 3 | 3 | 3 | 3 | 3 | 3 | 2 | 3 | 3 | 3 | 3 |
| Collinder 350 | 2 | 2 | 2 | 1 | 2 | 2 | 2 | 2 | 2 | 2 | 2 | 2 | 2 | 2 | 2 | 2 | 2 | 2 | 0 | 1 | 2 | 2 | 2 |
| Presepe-NGC 2632 | 1 | 1 | 1 | 0 | 1 | 1 | 1 | 1 | 1 | 1 | 1 | 1 | 1 | 1 | 1 | 1 | 1 | 1 | 0 | 1 | 1 | 1 | 1 |
| UBC 60 | 1 | 1 | 1 | 0 | 1 | 1 | 1 | 1 | 1 | 1 | 1 | 1 | 1 | 1 | 1 | 1 | 1 | 1 | 0 | 1 | 1 | 1 | 1 |
| UPK 84 | 3 | 3 | 3 | 0 | 3 | 3 | 3 | 3 | 3 | 3 | 3 | 3 | 3 | 3 | 3 | 3 | 3 | 3 | 1 | 0 | 3 | 2 | 3 |
| Theia 1214/NGC 752 | 6 | 6 | 6 | 0 | 6 | 6 | 6 | 6 | 6 | 6 | 6 | 6 | 6 | 6 | 6 | 6 | 6 | 6 | 2 | 2 | 6 | 6 | 6 |
| IC 4756 | 9 | 9 | 9 | 0 | 9 | 9 | 9 | 9 | 9 | 9 | 9 | 9 | 7 | 9 | 9 | 9 | 9 | 9 | 2 | 1 | 9 | 9 | 9 |
| Alessi 1 | 4 | 4 | 4 | 0 | 4 | 4 | 4 | 4 | 4 | 4 | 4 | 4 | 4 | 4 | 4 | 4 | 4 | 4 | 3 | 2 | 4 | 4 | 4 |
| NGC 2509 | 1 | 1 | 1 | 0 | 1 | 1 | 1 | 0 | 1 | 1 | 0 | 1 | 1 | 1 | 1 | 1 | 1 | 1 | 0 | 0 | 1 | 0 | 0 |
| NGC 7789 | 3 | 3 | 3 | 0 | 3 | 3 | 3 | 3 | 3 | 3 | 3 | 3 | 3 | 3 | 3 | 3 | 3 | 3 | 2 | 3 | 3 | 3 | 3 |
| NGC 7044 | 4 | 4 | 4 | 0 | 4 | 4 | 4 | 4 | 4 | 4 | 4 | 4 | 4 | 4 | 4 | 4 | 4 | 4 | 3 | 4 | 4 | 4 | 4 |
| NGC 6939 | 4 | 4 | 4 | 0 | 4 | 4 | 4 | 4 | 4 | 4 | 4 | 4 | 4 | 4 | 4 | 4 | 4 | 4 | 4 | 3 | 4 | 4 | 4 |
| NGC 2420 | 1 | 1 | 1 | 0 | 1 | 1 | 1 | 1 | 1 | 1 | 1 | 1 | 1 | 1 | 1 | 1 | 1 | 1 | 1 | 1 | 1 | 1 | 1 |
| Collinder 110 | 3 | 3 | 3 | 0 | 3 | 3 | 3 | 3 | 3 | 3 | 3 | 3 | 3 | 3 | 3 | 3 | 3 | 3 | 3 | 3 | 3 | 3 | 3 |
| NGC 7762 | 2 | 2 | 2 | 0 | 2 | 2 | 2 | 2 | 2 | 2 | 2 | 2 | 2 | 2 | 2 | 2 | 2 | 2 | 2 | 1 | 2 | 2 | 2 |
| UBC 141 | 2 | 2 | 2 | 0 | 2 | 2 | 2 | 1 | 2 | 2 | 2 | 1 | 2 | 2 | 2 | 1 | 2 | 1 | 0 | 1 | 2 | 1 | 1 |
| NGC 6819 | 2 | 2 | 2 | 0 | 2 | 2 | 2 | 2 | 2 | 2 | 2 | 2 | 2 | 2 | 2 | 2 | 2 | 2 | 1 | 2 | 2 | 2 | 2 |
| Ruprecht 171 | 6 | 6 | 6 | 0 | 6 | 6 | 6 | 6 | 6 | 6 | 6 | 6 | 6 | 6 | 6 | 6 | 6 | 6 | 3 | 3 | 6 | 6 | 6 |
| UBC 577 | 3 | 3 | 3 | 0 | 3 | 3 | 3 | 3 | 3 | 3 | 3 | 3 | 2 | 3 | 3 | 3 | 3 | 3 | 0 | 1 | 3 | 3 | 2 |
| NGC 7142 | 3 | 3 | 3 | 0 | 3 | 3 | 3 | 3 | 3 | 3 | 3 | 3 | 3 | 3 | 3 | 3 | 3 | 3 | 3 | 3 | 3 | 3 | 3 |
| NGC 2683 | 1 | 1 | 1 | 0 | 1 | 1 | 1 | 1 | 1 | 1 | 1 | 1 | 1 | 1 | 1 | 1 | 1 | 1 | 0 | 0 | 1 | 1 | 1 |
| Trumpler 5 | 3 | 3 | 3 | 0 | 3 | 3 | 3 | 3 | 3 | 3 | 3 | 3 | 3 | 3 | 3 | 3 | 3 | 3 | 1 | 3 | 3 | 3 | 3 |
| King 11 | 5 | 5 | 5 | 0 | 5 | 5 | 5 | 5 | 5 | 5 | 5 | 5 | 5 | 5 | 5 | 5 | 5 | 5 | 4 | 4 | 5 | 5 | 4 |
| Berkeley 32 | 2 | 2 | 2 | 0 | 2 | 2 | 2 | 2 | 2 | 2 | 2 | 2 | 2 | 2 | 2 | 2 | 2 | 2 | 1 | 1 | 2 | 1 | 1 |
| NGC 6791 | 2 | 2 | 2 | 0 | 2 | 2 | 2 | 2 | 2 | 2 | 2 | 2 | 2 | 2 | 2 | 2 | 2 | 2 | 0 | 2 | 2 | 2 | 2 |
| Total | 114 | 114 | 114 | 2 | 114 | 114 | 114 | 112 | 113 | 114 | 112 | 114 | 110 | 114 | 114 | 114 | 113 | 114 | 53 | 67 | 112 | 111 | 106 |

Table D.2: [X/Fe] of all our individual sample stars. Only a segment of the table is presented here, and that the full table is available to access at the CDS.

| Stellar Cluster | Star | [Mg/Fe] (dex) | [Si/Fe] (dex) | [S/Fe] (dex) | [Ca/Fe] (dex) | [F/Fe] (dex) | [Na/Fe] (dex) | [Al/Fe] (dex) | [K/Fe] (dex) | [Ti/Fe] (dex) | [V/Fe] (dex) |
|---|---|---|---|---|---|---|---|---|---|---|---|
| Gulliver 18 | Gulliver_18 | 0.08 ± 0.02 | −0.04 ± 0.03 | 0.24 ± 0.03 | 0.05 ± 0.02 | 0.03 ± 0.02 | 0.18 ± 0.01 | −0.09 ± 0.04 | −0.01 ± 0.11 | −0.03 ± 0.04 | −0.05 ± 0.04 |
| Collinder 463 | Collinder_463_1 | 0.13 ± 0.03 | −0.05 ± 0.03 | 0.24 ± 0.02 | 0.05 ± 0.01 | – | 0.06 ± 0.02 | −0.08 ± 0.02 | −0.08 ± 0.03 | −0.04 ± 0.04 | −0.19 ± 0.04 |
| | Collinder_463_2 | 0.09 ± 0.02 | −0.03 ± 0.03 | 0.38 ± 0.01 | 0.08 ± 0.01 | – | 0.16 ± 0.02 | −0.04 ± 0.03 | 0.00 ± 0.03 | −0.02 ± 0.04 | −0.22 ± 0.06 |
| | Collinder_463_3 | 0.11 ± 0.03 | −0.06 ± 0.03 | 0.38 ± 0.02 | 0.09 ± 0.02 | – | 0.14 ± 0.02 | −0.06 ± 0.03 | 0.10 ± 0.03 | −0.08 ± 0.05 | −0.19 ± 0.04 |
| Alessi Teutsch 11 | Alessi_Teutsch_11 | 0.13 ± 0.03 | 0.08 ± 0.03 | 0.49 ± 0.04 | 0.13 ± 0.02 | – | 0.07 ± 0.05 | 0.06 ± 0.02 | 0.03 ± 0.01 | 0.07 ± 0.04 | 0.03 ± 0.03 |
| UPK 219 | UPK_219 | 0.07 ± 0.03 | −0.12 ± 0.03 | 0.13 ± 0.04 | 0.00 ± 0.01 | – | −0.08 ± 0.04 | −0.09 ± 0.02 | −0.17 ± 0.01 | −0.03 ± 0.05 | 0.09 ± 0.06 |
| Gulliver 24 | Gulliver_24 | 0.04 ± 0.03 | 0.11 ± 0.03 | 0.46 ± 0.08 | −0.08 ± 0.03 | – | 0.02 ± 0.04 | −0.06 ± 0.02 | −0.12 ± 0.06 | −0.06 ± 0.05 | −0.06 ± 0.03 |
| Tombaugh 5 | Tombaugh_5_2 | 0.08 ± 0.03 | −0.03 ± 0.04 | 0.44 ± 0.08 | 0.12 ± 0.02 | – | 0.27 ± 0.04 | −0.05 ± 0.04 | −0.28 ± 0.10 | −0.07 ± 0.06 | −0.27 ± 0.10 |
| | Tombaugh_5_4 | 0.11 ± 0.02 | 0.00 ± 0.03 | 0.25 ± 0.02 | 0.14 ± 0.01 | – | 0.18 ± 0.03 | −0.01 ± 0.02 | −0.02 ± 0.02 | 0.04 ± 0.04 | −0.11 ± 0.07 |
| ⋮ | ⋮ | ⋮ | ⋮ | ⋮ | ⋮ | ⋮ | ⋮ | ⋮ | ⋮ | ⋮ | ⋮ |
| ⋮ | ⋮ | ⋮ | ⋮ | ⋮ | ⋮ | ⋮ | ⋮ | ⋮ | ⋮ | ⋮ | ⋮ |
| ⋮ | ⋮ | ⋮ | ⋮ | ⋮ | ⋮ | ⋮ | ⋮ | ⋮ | ⋮ | ⋮ | ⋮ |

| Stellar Cluster | Star | [Cr/Fe] (dex) | [Mn/Fe] (dex) | [Co/Fe] (dex) | [Ni/Fe] (dex) | [Cu/Fe] (dex) | [Zn/Fe] (dex) | [Y/Fe] (dex) | [Ce/Fe] (dex) | [Nd/Fe] (dex) | [Yb/Fe] (dex) |
|---|---|---|---|---|---|---|---|---|---|---|---|
| Gulliver 18 | Gulliver_18 | $-0.09 \pm 0.08$ | $0.14 \pm 0.02$ | $-0.03 \pm 0.01$ | $0.06 \pm 0.02$ | $0.09 \pm 0.04$ | - | $-0.07 \pm 0.03$ | $0.24 \pm 0.02$ | $0.30 \pm 0.06$ | $0.30 \pm 0.06$ |
| | Collinder_463_1 | $-0.08 \pm 0.02$ | $0.08 \pm 0.02$ | $-0.09 \pm 0.01$ | $0.11 \pm 0.01$ | $-0.07 \pm 0.06$ | - | - | $0.19 \pm 0.03$ | $0.19 \pm 0.07$ | $0.49 \pm 0.01$ |
| Collinder 463 | Collinder_463_2 | $-0.07 \pm 0.03$ | $0.04 \pm 0.02$ | $-0.09 \pm 0.01$ | $0.11 \pm 0.02$ | $-0.16 \pm 0.13$ | - | $-0.00 \pm 0.23$ | $0.21 \pm 0.02$ | $0.28 \pm 0.04$ | $0.53 \pm 0.03$ |
| | Collinder_463_3 | $-0.10 \pm 0.03$ | $0.07 \pm 0.03$ | $-0.11 \pm 0.01$ | $0.10 \pm 0.02$ | $0.05 \pm 0.06$ | - | - | $0.15 \pm 0.03$ | $0.17 \pm 0.13$ | $0.56 \pm 0.02$ |
| Alessi Teutsch 11 | Alessi_Teutsch_11 | $-0.16 \pm 0.05$ | $0.09 \pm 0.01$ | $-0.05 \pm 0.01$ | $0.15 \pm 0.01$ | $-0.11 \pm 0.06$ | $0.44 \pm 0.11$ | $-0.13 \pm 0.06$ | $0.30 \pm 0.03$ | $0.34 \pm 0.07$ | $0.58 \pm 0.03$ |
| UPK 219 | UPK_219 | $-0.10 \pm 0.03$ | $-0.07 \pm 0.03$ | $-0.18 \pm 0.01$ | $0.01 \pm 0.02$ | $-0.09 \pm 0.04$ | - | - | $0.16 \pm 0.04$ | $0.34 \pm 0.05$ | $0.32 \pm 0.02$ |
| Gulliver 24 | Gulliver_24 | $-0.05 \pm 0.04$ | $0.07 \pm 0.05$ | $-0.01 \pm 0.01$ | $0.13 \pm 0.01$ | $-0.18 \pm 0.05$ | $0.50 \pm 0.10$ | $-0.12 \pm 0.08$ | $0.45 \pm 0.02$ | $0.45 \pm 0.06$ | $0.75 \pm 0.04$ |
| Tombaugh 5 | Tombaugh_5_2 | $-0.03 \pm 0.04$ | $-0.04 \pm 0.09$ | $-0.08 \pm 0.03$ | $0.13 \pm 0.03$ | $0.17 \pm 0.08$ | - | - | $0.27 \pm 0.05$ | $0.32 \pm 0.11$ | $-0.32 \pm 0.13$ |
| | Tombaugh_5_4 | $-0.12 \pm 0.12$ | $-0.01 \pm 0.08$ | $-0.05 \pm 0.02$ | $0.13 \pm 0.02$ | $0.03 \pm 0.11$ | $-0.58 \pm 0.60$ | $-0.24 \pm 0.32$ | $0.26 \pm 0.04$ | $0.25 \pm 0.06$ | $0.52 \pm 0.03$ |
| ... | ... | ... | ... | ... | ... | ... | ... | ... | ... | ... | ... |
| ... | ... | ... | ... | ... | ... | ... | ... | ... | ... | ... | ... |
| ... | ... | ... | ... | ... | ... | ... | ... | ... | ... | ... | ... |

Table D.3: Similar to the table D.2 but for the remaining elements.

Table D.4: Mean [Fe/H], [Mg/Fe], [Si/Fe], [S/Fe], [Ca/Fe], [F/Fe], [Na/Fe], [Al/Fe], and [K/Fe] for the clusters in our sample.

| Stellar Cluster | [Fe/H] (dex) | [Mg/Fe] (dex) | [Si/Fe] (dex) | [S/Fe] (dex) | [Ca/Fe] (dex) | [F/Fe] (dex) | [Na/Fe] (dex) | [Al/Fe] (dex) | [K/Fe] (dex) |
|---|---|---|---|---|---|---|---|---|---|
| Gulliver 18 | -0.03 ± 0.03 | 0.08 ± 0.03 | -0.04 ± 0.03 | 0.24 ± 0.10 | 0.05 ± 0.04 | 0.03 ± 0.01 | 0.18 ± 0.04 | -0.09 ± 0.04 | -0.01 ± 0.11 |
| Collinder 463 | 0.03 ± 0.02 | 0.11 ± 0.02 | -0.05 ± 0.01 | 0.33 ± 0.08 | 0.07 ± 0.02 | - | 0.12 ± 0.06 | -0.06 ± 0.02 | 0.01 ± 0.09 |
| Alessi Teutsch 11 | -0.10 ± 0.03 | 0.13 ± 0.03 | 0.08 ± 0.03 | 0.49 ± 0.10 | 0.13 ± 0.04 | - | 0.07 ± 0.07 | 0.06 ± 0.04 | 0.03 ± 0.11 |
| UPK 219 | 0.06 ± 0.03 | 0.07 ± 0.03 | -0.12 ± 0.03 | 0.13 ± 0.10 | 0.00 ± 0.04 | - | -0.08 ± 0.04 | -0.09 ± 0.04 | -0.17 ± 0.10 |
| Gulliver 24 | -0.05 ± 0.03 | 0.04 ± 0.03 | 0.11 ± 0.03 | 0.46 ± 0.10 | -0.08 ± 0.04 | - | 0.02 ± 0.06 | -0.06 ± 0.04 | -0.12 ± 0.10 |
| Tombaugh 5 | 0.03 ± 0.00 | 0.10 ± 0.02 | -0.02 ± 0.03 | 0.35 ± 0.10 | 0.13 ± 0.01 | - | 0.22 ± 0.06 | -0.03 ± 0.03 | -0.15 ± 0.18 |
| NGC 7086 | -0.13 ± 0.03 | 0.16 ± 0.03 | 0.11 ± 0.03 | 0.40 ± 0.10 | 0.05 ± 0.04 | - | -0.03 ± 0.07 | -0.02 ± 0.03 | -0.11 ± 0.11 |
| Basel 11b | -0.03 ± 0.01 | 0.11 ± 0.02 | 0.04 ± 0.01 | 0.31 ± 0.13 | 0.07 ± 0.03 | - | 0.08 ± 0.07 | -0.02 ± 0.03 | 0.05 ± 0.13 |
| UBC 194 | -0.04 ± 0.03 | 0.06 ± 0.03 | 0.02 ± 0.03 | 0.25 ± 0.10 | 0.09 ± 0.04 | - | 0.15 ± 0.07 | 0.06 ± 0.04 | 0.06 ± 0.13 |
| COIN-Gaia 30 | -0.13 ± 0.03 | 0.14 ± 0.03 | 0.08 ± 0.03 | 0.37 ± 0.10 | 0.14 ± 0.04 | - | 0.15 ± 0.06 | -0.01 ± 0.04 | -0.07 ± 0.12 |
| UBC 169 | -0.02 ± 0.08 | 0.10 ± 0.04 | 0.02 ± 0.01 | 0.10 ± 0.10 | 0.11 ± 0.04 | - | 0.16 ± 0.06 | 0.08 ± 0.02 | -0.07 ± 0.12 |
| NGC 2437 | 0.00 ± 0.01 | 0.10 ± 0.01 | -0.01 ± 0.03 | 0.22 ± 0.09 | 0.09 ± 0.04 | - | 0.06 ± 0.24 | 0.02 ± 0.04 | 0.08 ± 0.07 |
| LP 1800 | -0.01 ± 0.02 | 0.10 ± 0.03 | 0.05 ± 0.01 | 0.28 ± 0.02 | 0.04 ± 0.06 | - | 0.25 ± 0.14 | 0.06 ± 0.09 | 0.07 ± 0.13 |
| Gulliver 51 | -0.09 ± 0.03 | 0.09 ± 0.03 | 0.04 ± 0.02 | 0.29 ± 0.10 | 0.15 ± 0.04 | - | 0.02 ± 0.03 | 0.02 ± 0.02 | - |
| Stock 2 | 0.01 ± 0.04 | 0.09 ± 0.03 | -0.07 ± 0.02 | 0.18 ± 0.04 | 0.06 ± 0.05 | - | 0.07 ± 0.07 | -0.06 ± 0.06 | -0.01 ± 0.07 |
| NGC 2548 | 0.01 ± 0.04 | 0.08 ± 0.03 | -0.04 ± 0.02 | 0.24 ± 0.07 | 0.09 ± 0.03 | - | 0.14 ± 0.12 | -0.04 ± 0.07 | -0.02 ± 0.14 |
| NGC 7209 | -0.01 ± 0.04 | 0.07 ± 0.05 | 0.04 ± 0.01 | 0.41 ± 0.22 | 0.05 ± 0.02 | - | 0.38 ± 0.14 | 0.08 ± 0.04 | -0.06 ± 0.22 |
| Collinder 350 | 0.01 ± 0.10 | 0.03 ± 0.02 | -0.06 ± 0.02 | 0.08 ± 0.11 | 0.00 ± 0.04 | -0.02 ± 0.01 | 0.03 ± 0.01 | -0.07 ± 0.04 | -0.10 ± 0.14 |
| Presepe-NGC 2632 | 0.18 ± 0.01 | 0.12 ± 0.03 | -0.09 ± 0.03 | 0.09 ± 0.05 | 0.07 ± 0.02 | - | 0.16 ± 0.12 | -0.05 ± 0.04 | -0.03 ± 0.14 |
| UBC 60 | 0.13 ± 0.02 | 0.10 ± 0.02 | -0.03 ± 0.00 | 0.24 ± 0.05 | 0.05 ± 0.04 | - | 0.17 ± 0.06 | -0.02 ± 0.03 | -0.03 ± 0.11 |
| UPK 84 | -0.06 ± 0.03 | 0.10 ± 0.03 | -0.01 ± 0.03 | 0.13 ± 0.10 | 0.11 ± 0.05 | - | 0.02 ± 0.07 | -0.03 ± 0.02 | 0.00 ± 0.13 |
| Theia 1214/NGC 752 | -0.05 ± 0.04 | 0.10 ± 0.05 | -0.02 ± 0.03 | 0.19 ± 0.08 | 0.03 ± 0.04 | - | 0.00 ± 0.04 | -0.03 ± 0.04 | -0.04 ± 0.14 |
| IC 4756 | -0.05 ± 0.04 | 0.07 ± 0.03 | -0.04 ± 0.03 | 0.16 ± 0.11 | 0.01 ± 0.04 | - | 0.14 ± 0.15 | -0.03 ± 0.03 | 0.02 ± 0.14 |
| Alessi 1 | 0.01 ± 0.04 | 0.08 ± 0.03 | -0.03 ± 0.02 | 0.10 ± 0.07 | 0.05 ± 0.04 | - | 0.07 ± 0.08 | -0.02 ± 0.03 | -0.03 ± 0.12 |
| NGC 2509 | 0.02 ± 0.04 | 0.08 ± 0.03 | 0.10 ± 0.08 | 0.04 ± 0.11 | - | - | - | - | - |
| NGC 7789 | -0.02 ± 0.08 | 0.07 ± 0.05 | -0.06 ± 0.04 | 0.41 ± 0.24 | 0.10 ± 0.03 | - | 0.09 ± 0.12 | 0.11 ± 0.03 | -0.03 ± 0.14 |
| NGC 7044 | 0.03 ± 0.03 | 0.06 ± 0.03 | -0.01 ± 0.01 | 0.34 ± 0.10 | 0.08 ± 0.04 | - | 0.09 ± 0.07 | 0.02 ± 0.08 | -0.18 ± 0.01 |
| NGC 6939 | -0.02 ± 0.03 | 0.08 ± 0.02 | -0.03 ± 0.03 | 0.39 ± 0.14 | 0.05 ± 0.04 | - | 0.04 ± 0.07 | -0.04 ± 0.02 | -0.05 ± 0.10 |
| NGC 2420 | -0.22 ± 0.04 | 0.07 ± 0.05 | -0.01 ± 0.03 | 0.29 ± 0.24 | 0.06 ± 0.03 | - | -0.22 ± 0.12 | -0.06 ± 0.04 | -0.06 ± 0.01 |
| Collinder 110 | -0.18 ± 0.00 | 0.12 ± 0.05 | 0.02 ± 0.05 | 0.57 ± 0.16 | 0.03 ± 0.03 | - | -0.07 ± 0.06 | -0.01 ± 0.01 | -0.11 ± 0.01 |
| NGC 7762 | -0.06 ± 0.06 | 0.12 ± 0.09 | 0.01 ± 0.01 | 0.52 ± 0.23 | 0.11 ± 0.07 | - | 0.13 ± 0.07 | 0.04 ± 0.03 | -0.05 ± 0.03 |
| UBC 141 | -0.10 ± 0.05 | 0.10 ± 0.09 | -0.01 ± 0.03 | 0.43 ± 0.13 | 0.16 ± 0.07 | - | 0.08 ± 0.06 | 0.01 ± 0.08 | -0.03 ± 0.03 |
| NGC 6819 | -0.02 ± 0.09 | 0.01 ± 0.09 | -0.11 ± 0.03 | 0.29 ± 0.13 | 0.07 ± 0.04 | - | 0.08 ± 0.07 | 0.00 ± 0.03 | -0.05 ± 0.14 |
| Ruprecht 171 | -0.00 ± 0.03 | 0.11 ± 0.09 | 0.00 ± 0.00 | 0.22 ± 0.04 | 0.10 ± 0.04 | - | 0.06 ± 0.03 | 0.02 ± 0.04 | -0.01 ± 0.03 |
| UBC 577 | -0.09 ± 0.03 | 0.09 ± 0.09 | -0.03 ± 0.07 | 0.08 ± 0.10 | 0.01 ± 0.07 | - | 0.04 ± 0.07 | -0.03 ± 0.03 | 0.04 ± 0.11 |
| NGC 7142 | -0.03 ± 0.06 | 0.16 ± 0.09 | 0.01 ± 0.03 | 0.56 ± 0.13 | 0.09 ± 0.07 | - | 0.17 ± 0.05 | 0.12 ± 0.08 | -0.06 ± 0.03 |
| NGC 2682 | 0.01 ± 0.05 | 0.13 ± 0.03 | -0.07 ± 0.04 | 0.26 ± 0.23 | -0.01 ± 0.04 | - | 0.02 ± 0.13 | 0.00 ± 0.03 | -0.07 ± 0.05 |
| Trumpler 5 | -0.48 ± 0.06 | 0.17 ± 0.09 | 0.05 ± 0.02 | 0.56 ± 0.16 | 0.12 ± 0.07 | - | 0.08 ± 0.05 | 0.05 ± 0.08 | 0.10 ± 0.35 |
| King 11 | -0.27 ± 0.05 | 0.08 ± 0.09 | 0.07 ± 0.01 | 0.39 ± 0.13 | 0.12 ± 0.03 | - | 0.06 ± 0.14 | 0.02 ± 0.03 | -0.05 ± 0.07 |
| Berkeley 32 | -0.29 ± 0.09 | 0.14 ± 0.03 | 0.12 ± 0.02 | 0.39 ± 0.15 | 0.06 ± 0.05 | - | -0.02 ± 0.05 | -0.05 ± 0.08 | -0.11 ± 0.05 |
| NGC 6791 | 0.28 ± 0.03 | 0.12 ± 0.05 | -0.30 ± 0.09 | 0.44 ± 0.13 | 0.14 ± 0.06 | - | 0.33 ± 0.35 | 0.17 ± 0.08 | 0.45 ± 0.07 |

Table D.5: Mean [Ti/Fe], [V/Fe], [Cr/Fe], [Mn/Fe], [Co/Fe], and [Ni/Fe] for the clusters in our sample.

| Stellar Cluster | [Ti/Fe] (dex) | [V/Fe] (dex) | [Cr/Fe] (dex) | [Mn/Fe] (dex) | [Co/Fe] (dex) | [Ni/Fe] (dex) |
|---|---|---|---|---|---|---|
| Gulliver 18 | -0.03 ± 0.04 | -0.05 ± 0.16 | -0.09 ± 0.05 | 0.14 ± 0.05 | -0.03 ± 0.04 | 0.06 ± 0.04 |
| Collinder 463 | -0.04 ± 0.02 | -0.20 ± 0.02 | -0.08 ± 0.02 | 0.07 ± 0.02 | -0.10 ± 0.01 | 0.11 ± 0.01 |
| Alessi Teutsch 11 | 0.07 ± 0.04 | 0.03 ± 0.16 | -0.16 ± 0.05 | 0.09 ± 0.05 | -0.05 ± 0.03 | 0.15 ± 0.07 |
| UPK 219 | -0.03 ± 0.04 | 0.09 ± 0.26 | -0.10 ± 0.05 | -0.07 ± 0.02 | -0.18 ± 0.04 | 0.01 ± 0.07 |
| Gulliver 24 | -0.06 ± 0.04 | -0.06 ± 0.16 | -0.04 ± 0.05 | 0.07 ± 0.04 | -0.02 ± 0.04 | 0.13 ± 0.07 |
| Tombaugh 5 | -0.01 ± 0.06 | -0.19 ± 0.11 | -0.07 ± 0.06 | -0.03 ± 0.01 | -0.07 ± 0.04 | 0.13 ± 0.07 |
| NGC 7086 | -0.05 ± 0.04 | -0.08 ± 0.26 | -0.27 ± 0.05 | 0.06 ± 0.05 | -0.07 ± 0.03 | 0.08 ± 0.07 |
| Basel 11b | -0.02 ± 0.00 | -0.26 ± 0.26 | -0.06 ± 0.04 | -0.05 ± 0.04 | -0.11 ± 0.04 | 0.06 ± 0.10 |
| UBC 194 | 0.10 ± 0.04 | -0.00 ± 0.16 | -0.09 ± 0.05 | 0.06 ± 0.05 | -0.14 ± 0.03 | 0.05 ± 0.07 |
| COIN-Gaia 30 | 0.00 ± 0.04 | 0.06 ± 0.16 | -0.15 ± 0.07 | -0.02 ± 0.04 | -0.16 ± 0.03 | 0.07 ± 0.07 |
| UBC 169 | 0.08 ± 0.01 | -0.11 ± 0.21 | 0.04 ± 0.07 | 0.04 ± 0.07 | 0.00 ± 0.04 | 0.13 ± 0.08 |
| NGC 2437 | 0.03 ± 0.03 | -0.17 ± 0.04 | -0.03 ± 0.04 | -0.04 ± 0.01 | -0.15 ± 0.05 | 0.05 ± 0.07 |
| LP 1800 | 0.02 ± 0.06 | -0.05 ± 0.25 | -0.04 ± 0.04 | 0.12 ± 0.05 | -0.10 ± 0.04 | 0.12 ± 0.10 |
| Gulliver 51 | -0.03 ± 0.04 | 0.19 ± 0.19 | -0.07 ± 0.04 | -0.07 ± 0.04 | -0.12 ± 0.03 | 0.12 ± 0.01 |
| Stock 2 | -0.09 ± 0.06 | -0.26 ± 0.01 | -0.03 ± 0.02 | -0.02 ± 0.01 | -0.13 ± 0.01 | 0.10 ± 0.01 |
| NGC 2548 | -0.03 ± 0.06 | -0.35 ± 0.01 | -0.00 ± 0.01 | -0.04 ± 0.01 | -0.15 ± 0.01 | 0.14 ± 0.01 |
| NGC 7209 | 0.03 ± 0.03 | 0.10 ± 0.04 | 0.00 ± 0.06 | 0.11 ± 0.05 | -0.03 ± 0.06 | 0.17 ± 0.02 |
| Collinder 350 | -0.07 ± 0.05 | -0.29 ± 0.12 | -0.10 ± 0.06 | -0.12 ± 0.04 | -0.13 ± 0.06 | 0.06 ± 0.01 |
| Presepe-NGC 2632 | 0.00 ± 0.04 | -0.04 ± 0.04 | -0.02 ± 0.04 | -0.06 ± 0.05 | -0.10 ± 0.07 | 0.05 ± 0.05 |
| UBC 60 | -0.05 ± 0.07 | -0.40 ± 0.22 | -0.01 ± 0.03 | -0.03 ± 0.05 | -0.14 ± 0.07 | 0.14 ± 0.02 |
| UPK 84 | 0.01 ± 0.04 | -0.12 ± 0.22 | -0.02 ± 0.04 | -0.07 ± 0.04 | -0.13 ± 0.07 | 0.05 ± 0.02 |
| Theia 1214/NGC 752 | 0.01 ± 0.06 | -0.19 ± 0.09 | -0.01 ± 0.03 | -0.06 ± 0.05 | -0.11 ± 0.04 | 0.10 ± 0.02 |
| IC 4756 | 0.02 ± 0.04 | -0.22 ± 0.33 | 0.02 ± 0.06 | -0.01 ± 0.07 | -0.10 ± 0.05 | 0.08 ± 0.03 |
| Alessi 1 | -0.01 ± 0.04 | -0.24 ± 0.21 | -0.02 ± 0.02 | -0.10 ± 0.01 | -0.11 ± 0.05 | 0.02 ± 0.02 |
| NGC 2509 | -0.18 ± 0.04 | 0.12 ± 0.16 | -0.02 ± 0.05 | -0.12 ± 0.05 | -0.04 ± 0.04 | -0.18 ± 0.05 |
| NGC 7789 | -0.05 ± 0.07 | 0.03 ± 0.03 | -0.04 ± 0.04 | 0.04 ± 0.12 | 0.02 ± 0.07 | 0.14 ± 0.03 |
| NGC 7044 | -0.05 ± 0.06 | 0.04 ± 0.07 | -0.09 ± 0.04 | 0.16 ± 0.04 | -0.03 ± 0.03 | 0.11 ± 0.01 |
| NGC 6939 | -0.04 ± 0.03 | -0.07 ± 0.21 | -0.05 ± 0.04 | 0.05 ± 0.03 | -0.04 ± 0.04 | 0.07 ± 0.09 |
| NGC 2420 | 0.03 ± 0.03 | -0.16 ± 0.16 | -0.10 ± 0.05 | -0.06 ± 0.05 | 0.00 ± 0.04 | -0.07 ± 0.05 |
| Collinder 110 | -0.03 ± 0.05 | -0.04 ± 0.03 | -0.12 ± 0.04 | 0.04 ± 0.02 | -0.02 ± 0.01 | 0.04 ± 0.01 |
| NGC 7762 | 0.07 ± 0.03 | 0.01 ± 0.07 | -0.02 ± 0.05 | 0.05 ± 0.01 | 0.03 ± 0.01 | 0.20 ± 0.03 |
| UBC 141 | 0.03 ± 0.01 | -0.17 ± 0.04 | -0.01 ± 0.04 | 0.03 ± 0.16 | 0.02 ± 0.02 | 0.23 ± 0.03 |
| NGC 6819 | -0.09 ± 0.04 | 0.17 ± 0.16 | -0.10 ± 0.02 | 0.09 ± 0.02 | -0.12 ± 0.06 | 0.12 ± 0.06 |
| Ruprecht 171 | 0.01 ± 0.06 | -0.12 ± 0.16 | -0.07 ± 0.06 | 0.01 ± 0.05 | -0.01 ± 0.09 | 0.14 ± 0.09 |
| UBC 577 | -0.09 ± 0.04 | -0.31 ± 0.16 | 0.00 ± 0.05 | -0.07 ± 0.11 | -0.10 ± 0.02 | 0.07 ± 0.04 |
| NGC 7142 | 0.04 ± 0.03 | 0.08 ± 0.30 | -0.05 ± 0.08 | 0.11 ± 0.04 | 0.02 ± 0.06 | 0.15 ± 0.05 |
| NGC 2682 | -0.10 ± 0.01 | 0.16 ± 0.07 | 0.06 ± 0.09 | -0.01 ± 0.05 | -0.13 ± 0.01 | 0.11 ± 0.14 |
| Trumpler 5 | 0.01 ± 0.01 | 0.01 ± 0.16 | -0.04 ± 0.11 | 0.02 ± 0.01 | 0.08 ± 0.11 | 0.12 ± 0.11 |
| King 11 | -0.06 ± 0.04 | 0.02 ± 0.28 | -0.11 ± 0.08 | -0.07 ± 0.03 | 0.03 ± 0.05 | 0.13 ± 0.14 |
| Berkeley 32 | 0.04 ± 0.01 | -0.10 ± 0.04 | -0.04 ± 0.05 | 0.09 ± 0.05 | 0.00 ± 0.06 | 0.09 ± 0.05 |
| NGC 6791 | 0.04 ± 0.01 | 0.78 ± 0.16 | -0.12 ± 0.11 | 0.28 ± 0.01 | 0.14 ± 0.11 | 0.37 ± 0.11 |

Table D.6: Mean [Cu/Fe], [Zn/Fe], [Y/Fe], [Ce/Fe], [Nd/Fe], and [Yb/Fe] for the clusters in our sample.

| Stellar Cluster | [Cu/Fe] (dex) | [Zn/Fe] (dex) | [Y/Fe] (dex) | [Ce/Fe] (dex) | [Nd/Fe] (dex) | [Yb/Fe] (dex) |
|---|---|---|---|---|---|---|
| Gulliver 18 | 0.09 ± 0.09 | - | -0.07 ± 0.12 | 0.24 ± 0.06 | 0.31 ± 0.11 | 0.30 ± 0.13 |
| Collinder 463 | -0.06 ± 0.11 | - | 0.00 ± 0.11 | 0.19 ± 0.02 | 0.22 ± 0.06 | 0.53 ± 0.04 |
| Alessi Teutsch 11 | -0.11 ± 0.09 | 0.44 ± 0.12 | -0.13 ± 0.06 | 0.30 ± 0.05 | 0.34 ± 0.11 | 0.59 ± 0.13 |
| UPK 219 | -0.09 ± 0.09 | - | 0.07±0.12 | 0.16 ± 0.01 | 0.34 ± 0.05 | 0.32 ± 0.13 |
| Gulliver 24 | -0.18 ± 0.09 | 0.50 ± 0.03 | -0.12 ± 0.06 | 0.46 ± 0.05 | 0.45 ± 0.12 | 0.75 ± 0.03 |
| Tombaugh 5 | 0.10 ± 0.10 | -0.58 ± 0.05 | -0.24 ± 0.06 | 0.27 ± 0.09 | 0.28 ± 0.18 | 0.52 ± 0.02 |
| NGC 7086 | -0.09 ± 0.09 | 0.27 ± 0.12 | 0.10 ± 0.06 | 0.24 ± 0.05 | 0.31 ± 0.12 | 0.52 ± 0.13 |
| Basel 11b | -0.12 ± 0.09 | - | 0.00 ± 0.13 | 0.18 ± 0.04 | 0.32 ± 0.13 | 0.42 ± 0.03 |
| UBC 194 | -0.17 ± 0.09 | 0.42 ± 0.02 | - | 0.11 ± 0.09 | 0.25 ± 0.12 | 0.42 ± 0.06 |
| COIN-Gaia 30 | 0.15 ± 0.09 | 0.14 ± 0.12 | -0.46 ± 0.13 | 0.30 ± 0.05 | 0.18 ± 0.03 | 0.47 ± 0.07 |
| UBC 169 | -0.17 ± 0.09 | 0.24 ± 0.12 | 0.30 ± 0.11 | 0.27 ± 0.09 | -0.08 ± 0.03 | 0.48 ± 0.06 |
| NGC 2437 | -0.12 ± 0.06 | -0.19 ± 0.09 | 0.05 ± 0.12 | 0.18 ± 0.05 | 0.37 ± 0.11 | 0.37 ± 0.06 |
| LP 1800 | -0.10 ± 0.12 | 0.33 ± 0.06 | 0.00 ± 0.09 | 0.29 ± 0.06 | 0.27 ± 0.08 | 0.54 ± 0.05 |
| Gulliver 51 | -0.19 ± 0.09 | 0.07 ± 0.05 | 0.01 ± 0.05 | 0.18 ± 0.09 | 0.20 ± 0.11 | 0.47 ± 0.10 |
| Stock 2 | 0.12 ± 0.06 | - | -0.10 ± 0.07 | 0.07 ± 0.06 | 0.25 ± 0.06 | 0.34 ± 0.06 |
| NGC 2548 | -0.11 ± 0.06 | - | 0.14 ± 0.05 | 0.07 ± 0.05 | 0.34 ± 0.08 | 0.46 ± 0.10 |
| NGC 7209 | -0.11 ± 0.06 | 0.38 ± 0.03 | -0.07 ± 0.11 | 0.23 ± 0.05 | 0.18 ± 0.11 | 0.47 ± 0.06 |
| Collinder 350 | -0.09 ± 0.10 | - | -0.15 ± 0.05 | 0.17 ± 0.05 | 0.33 ± 0.08 | 0.47 ± 0.07 |
| Presepe–NGC 2632 | -0.06 ± 0.10 | - | -0.27 ± 0.06 | -0.26 ± 0.05 | 0.21 ± 0.06 | 0.21 ± 0.06 |
| UBC 60 | -0.28 ± 0.06 | - | 0.06 ± 0.06 | -0.35 ± 0.03 | -0.02 ± 0.05 | 0.18 ± 0.06 |
| UPK 84 | -0.22 ± 0.06 | -0.03 ± 0.05 | - | 0.19 ± 0.05 | 0.31 ± 0.12 | 0.49 ± 0.07 |
| Theia 1214/NGC 752 | -0.08 ± 0.06 | 0.16 ± 0.12 | -0.26 ± 0.07 | 0.18 ± 0.05 | 0.30 ± 0.08 | 0.23 ± 0.06 |
| IC 4756 | -0.22 ± 0.06 | 0.12 ± 0.07 | -0.38 ± 0.06 | 0.14 ± 0.05 | 0.28 ± 0.10 | 0.25 ± 0.03 |
| Alessi 1 | -0.17 ± 0.06 | -0.31 ± 0.05 | 0.08 ± 0.11 | 0.19 ± 0.02 | 0.40 ± 0.10 | 0.36 ± 0.03 |
| NGC 2509 | -0.03 ± 0.06 | - | - | - | -0.06 ± 0.08 | - |
| NGC 7789 | -0.05 ± 0.06 | 0.03 ± 0.05 | -0.14 ± 0.19 | 0.20 ± 0.02 | 0.37 ± 0.07 | 0.78 ± 0.03 |
| NGC 7044 | -0.10 ± 0.06 | -0.05 ± 0.05 | -0.14 ± 0.06 | 0.13 ± 0.05 | 0.24 ± 0.11 | 0.47 ± 0.01 |
| NGC 6939 | -0.13 ± 0.06 | -0.05 ± 0.11 | -0.23 ± 0.05 | 0.16 ± 0.05 | 0.22 ± 0.09 | 0.43 ± 0.03 |
| NGC 2420 | 0.14 ± 0.06 | 0.53 ± 0.07 | -0.20 ± 0.13 | 0.24 ± 0.02 | 0.38 ± 0.08 | 0.59 ± 0.03 |
| Collinder 110 | -0.18 ± 0.09 | 0.23±0.12 | -0.06 ± 0.05 | 0.25 ± 0.03 | 0.40 ± 0.06 | 0.59 ± 0.01 |
| NGC 7762 | -0.13 ± 0.09 | -0.18±0.12 | -0.21 ± 0.05 | 0.24 ± 0.02 | 0.23 ± 0.07 | 0.44 ± 0.03 |
| UBC 141 | -0.09 ± 0.09 | -0.03 ± 0.06 | - | 0.40 ± 0.04 | 0.33 ± 0.06 | 0.49 ± 0.03 |
| NGC 6819 | -0.11 ± 0.06 | 0.15 ± 0.05 | -0.18 ± 0.02 | -0.07 ± 0.06 | 0.12 ± 0.03 | 0.19 ± 0.01 |
| Ruprecht 171 | -0.10 ± 0.06 | -0.22 ± 0.05 | -0.15 ± 0.19 | 0.10 ± 0.07 | 0.27 ± 0.06 | 0.38 ± 0.03 |
| UBC 577 | 0.05 ± 0.06 | - | -0.33 ± 0.02 | 0.26 ± 0.06 | 0.29 ± 0.03 | 0.53 ± 0.01 |
| NGC 7142 | -0.07 ± 0.06 | 0.06 ± 0.05 | -0.18 ± 0.04 | 0.08 ± 0.05 | 0.11 ± 0.04 | 0.13 ± 0.01 |
| NGC 2682 | -0.12 ± 0.06 | - | - | -0.16 ± 0.04 | 0.26 ± 0.07 | -0.17 ± 0.01 |
| Trumpler 5 | -0.18 ± 0.06 | -0.08 ± 0.05 | -0.23 ± 0.06 | 0.23 ± 0.03 | 0.33 ± 0.03 | 0.79 ± 0.01 |
| King 11 | -0.13 ± 0.06 | 0.33±0.18 | -0.25 ± 0.06 | 0.21 ± 0.02 | 0.32 ± 0.05 | 0.43 ± 0.01 |
| Berkeley 32 | -0.04 ± 0.06 | 0.31±0.12 | -0.01±0.05 | 0.32 ± 0.01 | 0.48 ± 0.01 | 0.50 ± 0.01 |
| NGC 6791 | -0.08 ± 0.06 | - | 0.07±0.04 | 0.01 ± 0.01 | 0.01 ± 0.01 | 0.52 ± 0.01 |